\documentclass{article}

\usepackage{fullpage}
\usepackage[utf8]{inputenc} 
\usepackage[T1]{fontenc}    
\usepackage{hyperref}       
\usepackage{url}            
\usepackage{booktabs}       
\usepackage{amsfonts}       
\usepackage{nicefrac}       
\usepackage{microtype}      
\usepackage{xcolor}         
\usepackage{amsmath,amsfonts,amssymb,amsthm}
\usepackage{thm-restate}
\usepackage{multirow}
\usepackage{graphicx}
\usepackage{subfig}
\usepackage{mathtools}

\newcommand{\argmax}{\mathrm{argmax}}
\newcommand{\eps}{\epsilon}

\newcommand{\comp}{{\text{Comp}}}
\newcommand{\concomp}{{\text{ConComp}}}
\newcommand{\View}{\mathrm{View}}	
\newcommand{\supp}{\mathrm{supp}}

\usepackage{xcolor}
\definecolor{DarkGreen}{rgb}{0.1,0.5,0.1}

\newcommand{\wz}[1]{\textcolor{purple}{[Wanrong: #1]}}

\usepackage{algorithmicx,algpseudocode, algorithm}

\newtheorem{theorem}{Theorem}
\numberwithin{theorem}{section}
\newtheorem{lemma}[theorem]{Lemma}
\numberwithin{lemma}{section}

\numberwithin{corollary}{section}
\newtheorem{definition}{Definition}
\numberwithin{definition}{section}

\numberwithin{assumption}{section}

\numberwithin{example}{section}
\newtheorem{proposition}[definition]{Proposition}
\numberwithin{proposition}{section}
\newtheorem{claim}[theorem]{Claim}
\numberwithin{claim}{section}

\title{Concurrent Composition Theorems for Differential Privacy}

\author{Salil Vadhan\thanks{Harvard John A. Paulson School of Engineering and Applied Sciences. Supported by a grant from the Sloan Foundation and a Simons Investigator Award. } \and Wanrong Zhang\thanks{Harvard John A. Paulson School of Engineering and Applied Sciences. Supported by a Computing Innovation Fellowship from the Computing Research Association (CRA) and the Computing Community Consortium (CCC).}}

\begin{document}
	
	\date{}
	\maketitle
	
	\begin{abstract}
				
		We study the concurrent composition properties of interactive differentially private mechanisms, whereby an adversary can arbitrarily interleave its queries to the different mechanisms. We prove that all composition theorems for non-interactive differentially private mechanisms extend to the concurrent composition of interactive differentially private mechanisms, whenever differential privacy is measured using the hypothesis testing framework of $f$-DP, which captures standard $(\eps,\delta)$-DP as a special case. We prove the concurrent composition theorem by showing that every interactive $f$-DP mechanism can be simulated by interactive post-processing of a non-interactive $f$-DP mechanism.

In concurrent and independent work, Lyu~\cite{lyu2022composition} proves a similar result to ours for $(\eps,\delta)$-DP, as well as a concurrent composition theorem for R\'enyi DP. We also provide a simple proof of Lyu's concurrent composition theorem for R\'enyi DP.  Lyu leaves the general case of $f$-DP as an open problem, which we solve in this paper.

	\end{abstract}


\section{Introduction}

\subsection{Differential Privacy}

Differential privacy is a statistical notion of database privacy, which ensures that the output of an algorithm will still have approximately the same distribution if a single data entry were to be changed. Differential privacy can be defined in terms of a general database space $\mathcal{X}$, and a binary {\em neighboring} relation on $\mathcal{X}$, which we think of as capturing whether ``two datasets'' differ on one individual's data.
For example, if databases are real-valued and contain a fixed number $n$ of entries, then $\mathcal{X} = \mathbb{R}^n$, and two datasets $x, x'\in \mathbb{R}^n$ are said to be \emph{neighboring} if they differ in at most one coordinate.

\begin{definition}[Differential Privacy \cite{DMNS06}]\label{def.dp}
	A randomized algorithm $\mathcal{M}: \mathcal{X} \rightarrow \mathcal{R}$ is \emph{$(\epsilon,\delta)$-differentially private} if for every pair of neighboring datasets $x, x'\in \mathcal{X}$, and for every subset of possible outputs $\mathcal{S} \subseteq \mathcal{R}$,
	$$\Pr[\mathcal{M}(x) \in \mathcal{S}] \leq \exp(\epsilon)\cdot \Pr[\mathcal{M}(x') \in \mathcal{S}] + \delta.$$
\end{definition}

Thus, differential privacy requires that for all neighboring datasets $x,x'$, $\mathcal{M}(x)$ and $\mathcal{M}(x') $ are close as probability distributions (as measured by the parameters $\eps$ and $\delta$). A number of variants of differential privacy have been defined based on other ways of measuring closeness, leading to Concentrated differential privacy (CDP) \cite{dwork2016concentrated,bun2016concentrated} and R\'enyi differential privacy (RDP) \cite{mironov2017renyi} and $f$-differential privacy ($f$-DP) \cite{dong2019gaussian}.

\subsection{Interactive Differential Privacy}

Definition \ref{def.dp} considers only non-interactive mechanisms $\mathcal{M}$  that release query answers in one shot, but data analysts often interact with a database in an adaptive fashion.  In fact, many useful primitives in differential privacy such as the Sparse Vector Technique \cite{DNRRV09,DNPR10, dwork2014algorithmic}, and the Private Multiplicative Weights \cite{hardt2010multiplicative} allow analysts to ask an adaptive sequence of queries about a dataset. It motivates the study of {\em interactive mechanisms} to capture full-featured privacy-preserving data analytics. Here, we view the mechanism $\mathcal{M}$ as a party in an interactive protocol, interacting with a (possibly adversarial) analyst. 

\begin{definition}[Interactive protocols]\label{def.protocol}
	An  {\em interactive protocol} $(A,B)$ is any pair of functions on tuples of binary strings. The {\em interaction} between $A$ with input $x_A$ and $B$ with input $x_B$ is the following random process (denoted $(A(x_A), B(x_B))$):
	\begin{enumerate}
		\item Uniformly choose random coins $r_A$ and $r_B$ for $A$ and $B$, respectively.
		\item Repeat the following for $i=0,1,\ldots$.\\
		(a) If $i$ is even, let $m_i=A(x_A, m_1, m_3, \ldots, m_{i-1}; r_A)$.\\
		(b) If $i$ is odd, let $m_i=B(x_B, m_0, m_2, \ldots, m_{i-1}; r_B)$.\\
		(c) If $m_{i}=\texttt{halt}$, then exit loop.
	\end{enumerate}
\end{definition}

The view of a party in an interactive protocol captures everything the party ``sees'' during the execution.

\begin{definition}[View of a party in an interactive protocol]
	Let $(A,B)$ be an interactive protocol. Let $r_A$ and $r_B$ be the random coins for $A$ and $B$, respectively. $A$'s view of $(A(x_A;r_A), B(x_B; r_B))$ is the tuple $\View_A ( A(x_A;r_A) \leftarrow B(x_B; r_B) ) = (r_A, x_A, m_1, m_3, \ldots)$ consisting of all the messages received by $A$ in the execution of the protocol together with the private input $x_A$ and random coins $r_A$. $B$'s view of $(A(x_A;r_A), B(x_B; r_B))$ is defined symmetrically.
\end{definition}

%
%

In the setting of differentially private mechanisms,  Party $A$ is the mechanism, where the input $x_A$ is the dataset. party $B$ is the adversary that does not have an input $x_B$.  Since we only care about the view of the adversary, we will drop the subscript and denote the view of the adversary as $\View ( B \leftrightarrow  \mathcal{M}(x) )$. With this notation, interactive differential privacy is defined by asking for the views of an adversary on any pair of neighboring datasets $\View ( B \leftrightarrow  \mathcal{M}(x) )$ and $\View ( B \leftrightarrow  \mathcal{M}(x') )$ satisfying the same $(\eps, \delta)$-closeness notion as in non-interactive differential privacy. 


\begin{definition}
	A randomized algorithm $\mathcal{M}$ is an  {\em $(\eps,\delta)$-differentially private interactive mechanism} if for every pair of neighboring datasets $X,X'\in\mathcal{X}$, every adversary algorithm $B\in \mathcal{B}$, and every subset of possible views $\mathcal{S} \subseteq \text{Range}(\View )$,  we have 
	
	$$\Pr[ \View ( B \leftrightarrow \mathcal{M}(x) ) \in \mathcal{S}] \leq \exp(\epsilon)\cdot \Pr[ \View (B \leftrightarrow  \mathcal{M}(x') )\in \mathcal{S}] + \delta.$$
	
\end{definition}

\subsection{Concurrent Composition}

A fundamental problem in differential privacy is studying how the privacy degrades under {\em composition} as more computations are performed on the same database. The composition property is particularly useful when we want to ask interactive queries on the same database, and it also allows us to design a complex differentially private algorithm by combining several building blocks. Formally, we define the composition of a sequence of non-interactive $k$ mechanisms $\mathcal{M}_1, \mathcal{M}_2, \ldots, \mathcal{M}_{k}$ as the non-interactive mechanism $\mathcal{M}=\comp (\mathcal{M}_1, \mathcal{M}_2, \ldots, \mathcal{M}_{k}) $ defined as
\begin{equation}
	\mathcal{M}(x):= (\mathcal{M}_1(x), \mathcal{M}_2(x), \ldots, \mathcal{M}_{k}(x)),
\end{equation}
where each mechanism $\mathcal{M}$ is executed using independent random coins. 


The composition of non-interactive mechanisms has been studied extensively in the literature.
The basic composition theorem \cite{dwork2006our} states that the privacy parameters add up linearly when composing private mechanisms. The advanced composition theorem \cite{dwork2010boosting} provides a tighter bound where the privacy parameter grows sublinearly under $k$-fold adaptive composition. Later, the optimal composition theorem \cite{kairouz2015composition, murtagh2016complexity} gives an exact characterization of the privacy guarantee under $k$-fold composition. The relaxations of differential privacy including zero-concentrated differential privacy (zCDP) \cite{dwork2016concentrated,bun2016concentrated}, R\'enyi differential privacy (RDP) \cite{mironov2017renyi}, and $f$-differential privacy ($f$-DP) \cite{dong2019gaussian} allows for tighter reasoning about composition. 
In the abovementioned work, some of them \cite{dwork2010boosting, murtagh2016complexity} are framed in a way that the adversary can adaptively choose the mechanisms $\mathcal{M}_1, \mathcal{M}_2,\ldots,\mathcal{M}_{k}$, and thus the adaptive composition can be viewed as an interactive mechanism. 

\begin{figure}[H]
	\centering
	\includegraphics[width=.5\textwidth]{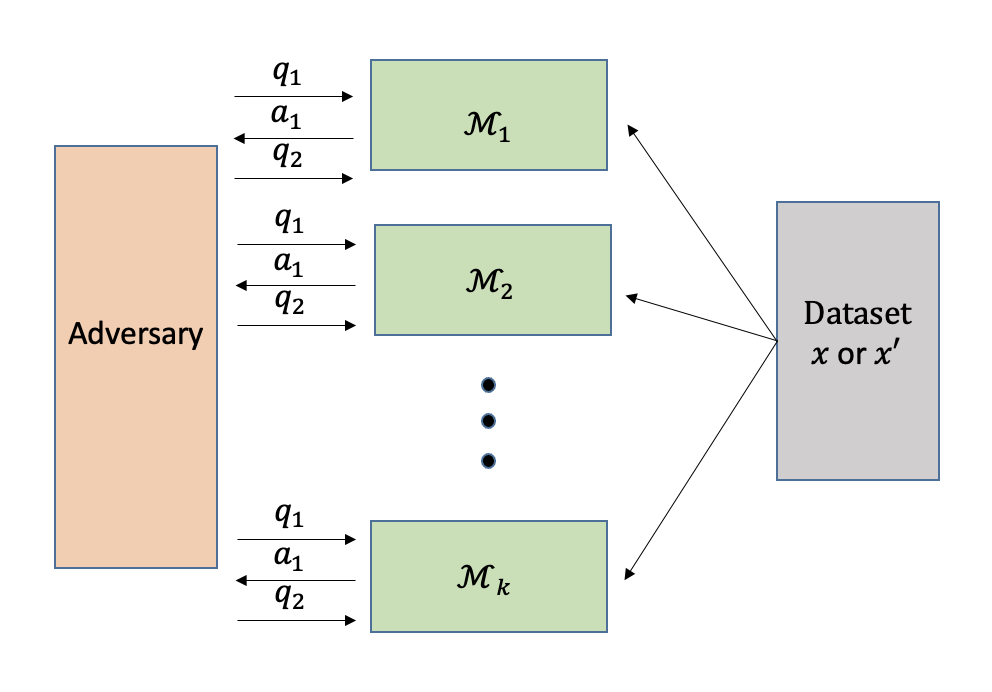}
	\caption{Concurrent composition of interactive mechanisms}\label{fig:concomp}
\end{figure}

In many cases, analysts may wish to perform multiple {\em interactive} analyses on the same dataset concurrently, which raises the question of {\em concurrent composition}, first studied for differential privacy in \cite{vadhan2021concurrent}.
In this setting (illustrated in Figure \ref{fig:concomp}), an adversary can arbitrarily interleave its queries to several differentially private mechanisms, and those queries might be correlated and depends on the answer received in other mechanisms. As a motivating example, several organizations might set up multiple DP query systems on datasets that may refer to the same set of individuals. Each query system has its own privacy budget $\eps$. Suppose an adversary can concurrently access those systems, and a query sent to one system might depends on all the previous messages that received from other systems. For example, when we run two Sparse Vector mechanisms $\mathcal{M}_1$ and $\mathcal{M}_2$ concurrently, the queries for $\mathcal{M}_1$ depends on previous answers from $\mathcal{M}_2$, and vice-versa, but we only know the overall privacy guarantees for $\mathcal{M}_1$ and $\mathcal{M}_2$ when they are executed independently. 
If the executions were {\em sequential}, meaning that the adversary completes its interaction with $\mathcal{M}_1$ before issuing any queries to $\mathcal{M}_2$, then we can hardwire the answers from   $\mathcal{M}_1$ into the adversary's strategy. When attacking $\mathcal{M}_2$, the privacy loss for $\mathcal{M}_2$ will be bounded as usual. But when the queries are interleaved, it is no longer clear how to define a fixed adversary strategy against either of the mechanisms $\mathcal{M}_1$ or $\mathcal{M}_2$.
It is not clear if the adversary can run any concurrent attack to break the privacy guarantees, and therefore, we wish to provide a provable guarantee to account for the total privacy loss in such systems. Formally, the concurrent composition of interactive mechanisms is defined as follows.

\begin{definition}[Concurrent composition of interactive mechanisms\cite{vadhan2021concurrent}]
	Let $\mathcal{M}_1,\ldots, \mathcal{M}_{k}$ be interactive mechanisms. $\mathcal{M}=ConComp(\mathcal{M}_1,\ldots, \mathcal{M}_{k})$ is the concurrent composition of mechanisms $\mathcal{M}_1,\ldots, \mathcal{M}_{k}$ defined as follows:
	\begin{enumerate}
		\item Random sample $r=(r_1,\ldots,r_{k})$ where $r_j$ are random coin tosses for $\mathcal{M}_j$.
		\item Inputs for $\mathcal{M}$ consists of $x=(x_1,\ldots,x_{k})$ where $x_j$ is a private dataset for $\mathcal{M}_j$. 
		\item $\mathcal{M}(x,m_0,\ldots,m_{i-1};r)$ is defined as follows:
		\begin{enumerate}
			\item Parse $m_{i-1}$ as $(j,q)$ where $j=1, \ldots, k$ and $q$ is a query to $\mathcal{M}_j$. If $m_{i-1}$ cannot be parsed correctly, output $\texttt{halt}$.
			\item Extract history $(m_0^j,\ldots,m_{t-1}^j)$ from $(m_0,\ldots,m_{i-1})$ where $m_i^j$ are all of the queries to mechanism $\mathcal{M}_j$. 
			\item Output $\mathcal{M}_j(x_j, m_0^j,\ldots,m_{t-1}^j; r_j)$.
		\end{enumerate}
	\end{enumerate}
For an adversary $B$, we will use the notation $\View(B\leftrightarrow (\mathcal{M}_1,\ldots, \mathcal{M}_{k}))$ as shorthand for $\View(B\leftrightarrow \concomp(\mathcal{M}_1,\ldots, \mathcal{M}_{k}))$
\end{definition}


Vadhan and Wang \cite{vadhan2021concurrent} showed that the advanced and optimal composition theorems extend to the concurrent composition of interactive pure DP mechanisms.

\begin{theorem}[\cite{vadhan2021concurrent}]\label{thm.puredpconcomp}
	Suppose that for all non-interactive mechanisms $\mathcal{M}_1, \ldots, \mathcal{M}_k$ such that $\mathcal{M}_i$ is $(\eps_i, \delta_i)$-differentially private for $\delta_1=\delta_2=\ldots=\delta_k=0$, their composition $\comp(\mathcal{M}_1, \ldots, \mathcal{M}_k)$ is $(\eps, \delta)$-differentially private. Then for all interactive mechanisms $\mathcal{M}_1, \ldots, \mathcal{M}_k$ such that $\mathcal{M}_i$ is $(\eps_i, \delta_i)$-differentially private for $\delta_1=\delta_2=\ldots=\delta_k=0$, 
	the concurrent composition $\concomp(\mathcal{M}_1, \ldots, \mathcal{M}_k)$ of interactive mechanisms $\mathcal{M}_1, \ldots, \mathcal{M}_k$ is $(\eps, \delta)$-differentially private.
\end{theorem}

They prove this by reducing the analysis of interactive pure DP mechanism to that of analyzing the Randomized Response mechanism \cite{warner1965randomized,DMNS06}:

\begin{theorem}[\cite{vadhan2021concurrent}]\label{thm.puredptoRR}
	Suppose that $\mathcal{M}$ is an interactive $(\eps, 0)$-differentially private mechanism. Then for every pair of neighboring datasets $x, x'$, there exists an interactive post-processing function $\mathcal{P}$  such that for every adversary $B \in \mathcal{B}$, we have 
	\begin{align}
		\View(B \leftrightarrow  \mathcal{M}(x))\equiv \View(B \leftrightarrow  \mathcal{P}(RR_\eps(0)))
		&&
		\View(B \leftrightarrow  \mathcal{M}(x'))\equiv \View(B \leftrightarrow \mathcal{P}(RR_\eps(1))).
	\end{align}
\end{theorem}

Here $\mathcal{P}$ is an interactive post-processing function that depends on $\mathcal{M}$ and a fixed pair of neighboring datasets $x, x'$. It receives a single bit as an output of $RR_\eps(0)$ or $RR_\eps(1)$, and then interacts with the adversary $A$.

Note that Theorem \ref{thm.puredpconcomp} and Theorem \ref{thm.puredptoRR} do not apply to the case where the composed mechanisms $\mathcal{M}_i$ are $(\eps_i,\delta_i)$-DP for $\delta_i>0$. In this case, \cite{vadhan2021concurrent} only show a bound that is similar to the ``group privacy'' property of $(\eps,\delta)$-DP. In particular, if $\eps_1=\eps_2=\ldots=\eps_k=\eps$ and $\delta_1=\delta_2=\ldots=\delta_k=\delta$, they show 
that the concurrent composition $\concomp(\mathcal{M}_1, \mathcal{M}_2, \ldots, \mathcal{M}_{k})$ is $(k\eps, \frac{\exp(k\eps)-1}{\exp(\eps)-1}\delta)$-differentially private.
This is suboptimal even compared to basic composition.
It left as an open problem that if any composition theorems for non-interactive mechanisms can extend to all variants of DP interactive mechanisms.

\paragraph{Open Question.} Does Theorem \ref{thm.puredpconcomp} extend to other variants of DP (such as $(\eps_i,\delta_i)$-DP with $\delta_i>0$, R\'enyi DP, $f$-DP)?


\subsection{Our Results on Concurrent Composition}

In this paper, we close this gap and show that any composition theorems of non-interactive mechanisms also extend to the concurrent composition of interactive DP mechanisms for approximate DP. In particular, we show that Theorem \ref{thm.puredpconcomp} extends to the case that $\delta_i>0$:

\begin{theorem}[Concurrent composition for $(\eps, \delta)$-DP interactive mechanisms]\label{thm.concomp}
	Suppose that for all non-interactive mechanisms $\mathcal{M}_1, \ldots, \mathcal{M}_k$ such that $\mathcal{M}_i$ is $(\eps_i, \delta_i)$-differentially private for $i=1,2 \ldots, k$, their composition $\comp(\mathcal{M}_1, \ldots, \mathcal{M}_k)$ is $(\eps, \delta)$-differentially private. Then for all interactive mechanisms $\mathcal{M}_1, \ldots, \mathcal{M}_k$ with finite communication such that $\mathcal{M}_i$ is $(\eps_i, \delta_i)$-differentially private for $i=1,2 \ldots, k$, 
	the concurrent composition $\concomp(\mathcal{M}_1, \ldots, \mathcal{M}_k)$ of interactive mechanisms $\mathcal{M}_1, \ldots, \mathcal{M}_k$ is $(\eps, \delta)$-differentially private.
\end{theorem}

We also handle general $f$-DP as defined and discussed in the section below.

\begin{theorem}[Concurrent composition for $f$-DP interactive mechanisms]\label{thm.fdp} 
	Suppose that for all non-interactive mechanisms $\mathcal{M}_1, \ldots, \mathcal{M}_k$ such that $\mathcal{M}_i$ is $f_i$-DP for $i=1,2 \ldots, k$, their composition $\comp(\mathcal{M}_1, \ldots, \mathcal{M}_k)$ is $f$-DP. Then for all interactive mechanisms $\mathcal{M}_1, \ldots, \mathcal{M}_k$ such that $\mathcal{M}_i$ is $f_i$-DP for $i=1,2 \ldots, k$, 
	the concurrent composition $\concomp(\mathcal{M}_1, \ldots, \mathcal{M}_k)$ of interactive mechanisms $\mathcal{M}_1, \ldots, \mathcal{M}_k$ is $f$-DP.
\end{theorem}



Theorem \ref{thm.concomp} follows directly from Theorem \ref{thm.fdp} because $f$-DP defined below captures $(\eps, \delta)$-DP as a special case 
\cite{wasserman2010statistical, dong2019gaussian}. Interestingly, the generalization to $f$-DP is important for our proof, even if we only want to prove Theorem \ref{thm.concomp}.
 We explain the detailed proof technique in the section below. 

In summary, our results show that there is no extra privacy loss due to the concurrent access to multiple interactive mechanisms. We can now safely run multiple interactive differentially private algorithms in parallel, while allowing communication with all them during their executions.

\subsection{$f$-DP and Interactive vs. Noninteractive Hypothesis Testing}\label{sec.introfdp}

$f$-differential privacy ($f$-DP) \cite{dong2019gaussian} is a generalization of $(\eps,\delta)$-differential privacy based on the hypothesis testing interpretation of differential privacy. Differential privacy attemps to measure the difficulty of distinguishing two neighboring datasets based on the ouput of a mechanism. Specifically, an adversary considers the following hypothesis testing problem:
$$H_0: \text{the dataset is }x \quad \text{versus} \quad H_1: \text{the dataset is }x'.$$

Denote by $Y$ and $Y'$ the output distributions of $\mathcal{M}$ on the two neighboring datasets, namely $\mathcal{M}(x)$ and $\mathcal{M}(x')$. For a given rejection rule $\phi$, the type I error $\alpha_\phi=\mathrm{E}[\phi(Y)]$ is the probability of rejecting $H_0$ when $H_0$ is true, while the type II error $\beta_\phi=1-\mathrm{E}[\phi(Y')]$ is the probability of failing to reject $H_0$ when $H_1$ is true. A trade-off function serves as the optimal boundary of the achievable and unachievable regions of these errors. 

\begin{definition}[Trade-off function \cite{dong2019gaussian}]\label{tradeoff.def}
	For any two probability distributions $Y$ and $Y'$ on the same space, define the {\em trade-off function} $T(Y,Y'): [0,1]\rightarrow [0,1]$ as
	\begin{equation}\label{eq.tradeoff}
		T(Y,Y')(\alpha)=\inf\{\beta_\phi: \alpha_\phi\le \alpha\},
	\end{equation}
	where the infimum is taken over all (measurable) rejection rules $\phi$.
\end{definition}

Proposition \ref{def.tradeoffclass} gives the necessary and sufficient condition for $f$ to be a trade-off function.
\begin{proposition}[Class of trade-off functions \cite{dong2019gaussian}]\label{def.tradeoffclass}
	A function $f: [0,1]\rightarrow [0,1]$ is a trade-off function if and only if $f$ is convex, continuous, non-increasing, and $f(x)\le 1-x$ for $x\in [0,1]$.
\end{proposition}

$f$-DP allows the full trade-off between type I and type II errors in the simple hypothesis testing problem to be governed by a trade-off function $f$. A larger trade-off functions implies stronger privacy guarantees.

\begin{definition}[$f$-differential privacy \cite{dong2019gaussian}]\label{def.fdp}
	Let $f$ be a trade-off function. A mechanism $\mathcal{M}: \mathcal{X} \rightarrow \mathcal{R}$  is {\em $f$-differentially private} if for every pair of neighboring datasets $x,x' \in \mathcal{X}$, we have
	$$T(\mathcal{M}(x), \mathcal{M}(x'))\ge f.$$
\end{definition}

$(\eps, \delta)$-DP is a special case of $f$-DP, taking $f=f_{\eps,\delta}$, where $f_{\eps,\delta}=\max\{0, 1-\delta-\exp(\eps)\alpha, \exp(-\eps)(1-\delta-\alpha)\}$ \cite{wasserman2010statistical, dong2019gaussian}. 

To prove Theorem \ref{thm.fdp} (and hence Theorem \ref{thm.concomp}), we prove the following analogue of Theorem \ref{thm.puredptoRR}, showing that every interactive $f$-DP mechanism can be simulated by an interactive post-processing of a non-interactive mechanism.

\begin{theorem}\label{thm.reduction}
	For every trade-off function $f$, every interactive $f$-DP mechanism $\mathcal{M}$ with finite communication, and every pair of neighboring datasets $x, x'$, there exists a non-interactive $f$-DP mechanism $\mathcal{N}$ and an randomized interactive post-processing mechanism $\mathcal{P}$ such that for every adversary $B \in \mathcal{B}$, we have 
	\begin{align}
		\View(B \leftrightarrow  \mathcal{M}(x))\equiv \View(B \leftrightarrow  \mathcal{P}(\mathcal{N}(x)))
		&&
		\View(B \leftrightarrow  \mathcal{M}(x'))\equiv \View(B \leftrightarrow  \mathcal{P}(\mathcal{N}(x'))).
	\end{align}
\end{theorem}
Similarly to Theorem \ref{thm.puredptoRR}, in the case of $(\eps,\delta)$-DP, one can take the non-interactive mechanism $\mathcal{N}$ as the $(\eps,\delta)$-Randomized Response mechanism of \cite{kairouz2015composition}.
Indeed, \cite{kairouz2015composition} shows that every non-interactive $(\eps,\delta)$-DP mechanism can be simulated as a post-processing of $(\eps,\delta)$-Randomized Response.

Theorem \ref{thm.fdp} follows from Theorem \ref{thm.reduction} in the same way as Theorem \ref{thm.puredpconcomp} follows from Theorem \ref{thm.puredptoRR}. Indeed, Theorem \ref{thm.fdp} implies that
to analyze the concurrent composition of interactive mechanisms $\mathcal{M}_i$, it suffices to consider the composition of the non-interactive mechanisms $\mathcal{N}_i$. As a result, composition theorems for non-interactive mechanisms extend to the concurrent composition of interactive $f$-DP  mechanisms. 

\newcommand{\cM}{\mathcal{M}}
\newcommand{\cN}{\mathcal{N}}

Theorem \ref{thm.reduction} is an interesting statement about statistical hypothesis testing even without the application to differential privacy.  Normally, hypothesis testing is presented as the task of distinguishing between two distributions or sets of distributions.  This is a noninteractive task: a sample from the distribution is generated and given to the hypothesis tester, which then tries to decide whether the distribution is in $H_0$ or $H_1$.  However, suppose instead we consider the task of distinguishing between two interactive mechanisms $\cM_0$ and $\cM_1$, each of which responds to queries in a randomized and stateful manner. Since the mechanisms are stateful, the hypothesis tester may never learn everything there is to know about the mechanism; in particular it cannot find out how the mechanism would have answered if different queries had been asked in the past.  This is in contrast to ordinary hypothesis testing, where the full sample from the distribution is given to the hypothesis tester.  Nevertheless, by viewing $\cM_0$ as $\cM(x)$ and $\cM_1$ as $\cM(x')$, Theorem~\ref{thm.reduction} implies that the two interactive mechanisms $\cM_0$ and $\cM_1$ can be simulated perfectly by noninteractive random variables $\cN_0=\cN(x)$ and $\cN_1=\cN(x')$ such that even if we give $\cN_0$ or $\cN_1$ to a hypothesis tester in its entirety (thereby revealing how $\cM_0$ or $\cM_1$ would answer {\em all} questions), it cannot distinguish them any better than it could distinguish $\cM_0$ and $\cM_1$.  The trick, of course, is that the simulation is ``perfect'' only when executing a single interaction with $\cM_0$ or $\cM_1$ (with no rewinding to explore multiple paths in the interaction tree).

The proof of Theorem \ref{thm.reduction} relies on the following two lemmas. 


\begin{lemma}[Coupling property of $f$-DP]\label{lemma.coupling}
	Let $f$ be a trade-off function, and suppose we have random variables $X$, $Y$  and $X'$, $Y'$ such that 
	\begin{equation*}
	T(X,X') \ge f \quad \text{and} \quad T(Y,Y')\ge f.
\end{equation*}
Then there exists couplings $(X,Y)$ and $(X',Y')$ such that
\begin{equation*}
	T((X,Y)||(X',Y'))\ge f.
\end{equation*}
\end{lemma}

A coupling of random variables $X$ and $Y$ is any random vector $(\tilde{X}, \tilde{Y})$ such that the marginal distributions are identically distributed to $X$ and $Y$ respectively, i.e., $\tilde{X} \equiv X$ and $\tilde{Y} \equiv Y$. Subject to this constraint on the marginal, $\tilde{X}$ and $\tilde{Y}$ can be arbitrarily correlated. Allowing correlations is critical to Lemma \ref{lemma.coupling}.
For example, for the case of $(\eps,\delta)$-DP, if we keep $X, Y$ and $X', Y'$ independent, then we would just get the ``group privacy'' like bound. 


\begin{lemma}[Chain rule of $f$-DP]\label{lemma.chain}
	For every pair of random variables $X, X'$ with finite support, there exists a function $\mathrm{ChainRule}_{X,X'}$ such that for every random variable $Y$ jointly distributed with $X$, and every random variable $Y'$ jointly distributed with $X'$, we have 
	\begin{equation*}
		T((X,Y), (X',Y'))=\mathrm{ChainRule}_{X,X'}((T(Y|X=x, Y'|X'=x)))_{x\in \supp(X)\cap \supp(X') }).
	\end{equation*}
	Moreover, $\mathrm{ChainRule}$ is a function that is ``continuous in each variable'' on the partially ordered set of trade-off functions (see Section \ref{sec.dDP} for formal definition).
\end{lemma}

Lemma \ref{lemma.chain} says that the trade-off function between $(X, Y)$ and $(X',Y')$ can be determined by a collection of trade-off functions between $Y$ and $Y'$ conditioned on $X=X'=x$ for every $x\in \supp(X)\cap \supp(X')$ through a $\mathrm{ChainRule}$ function.
The terminology ``chain rule'' is by analogy with the standard chain rule for KL divergence, which says

\begin{equation}\label{chain.KL}
\mathrm{KL}((X,Y)||(X',Y'))=\mathrm{KL}(X||X')+\mathrm{E}_{x\sim X}\mathrm{KL}(Y|X=x|| Y'|X'=x).
\end{equation}
So fixing $X$ and $X'$, we can calculate the KL divergence for arbitrary $Y$ and $Y'$ as a function of the KL divergences $\mathrm{KL}(Y|X=x|| Y'|X'=x)$.
 $(\eps,\delta)$-DP does not admit the chain rule property, because no pairs of $(\eps, \delta)$ can exactly capture the ``closeness'' of $(X,Y)$ and $(X',Y')$ given a collection of $\{\eps_j, \delta_j\}_j$ that characterize the ``closeness'' of $Y|X=x$ and $Y'|X'=x'$. Working with the general $f$-DP allows us to capture a complete characterization of ``privacy''.

To prove Theorem \ref{thm.reduction} using Lemmas \ref{lemma.coupling} and \ref{lemma.chain}, our strategy is to apply induction on  the number of messages exchanged (which we can do since $\mathcal{M}$ has finite communication by assumptions). To reduce $k$ rounds of interactions to $k-1$ rounds, we consider the subsequent interaction conditioned on the first message. Depending on whether the first message sent from the mechanism $\mathcal{M}$ or the adversary $B$, we consider the following two cases. 

\textbf{Case 1.} The adversary $B$ sends the first query $q_1$ to the mechanism $\mathcal{M}$. Fix a pair of neighboring datasets $x, x'$. Fixing $q_1$, we denote the subsequent interactive mechanism by $\mathcal{M}_{q_1}$. By induction, $\mathcal{M}_{q_1}$ can be simulated by a post-processing of a non-interactive $f$-DP mechanism $\mathcal{N}_{q_1}$. 
Then we obtain $\mathcal{N}(x)$ and $\mathcal{N}(x')$ by coupling the pairs $\mathcal{N}_{q_1}(x)$ and $\mathcal{N}_{q_1}(x')$ on all the values of $q_1$, which is finite by our assumption of finite communication. We note that the coupling lemma  \ref{lemma.coupling} extends to finite number of random variables $ Y_1, \ldots, Y_k$ and $Y'_1, \ldots, Y'_k$ by induction on $k$. Following the coupling lemma, we have $T(\mathcal{N}(x), \mathcal{N}(x') )\ge f$.
We can combine the interactive post-processing mechanisms $\mathcal{P}_{q_1}$ for all $q_1$ to obtain the interactive post-processing $\mathcal{P}$ that simulates $\mathcal{M}(x)$ and $\mathcal{M}(x')$ from $\mathcal{N}(x)$ and $\mathcal{N}(x')$. Thus, we have Theorem \ref{thm.reduction} holds for $k$ rounds of interactions. 

\begin{figure}[H]
	\centering
	\includegraphics[width=0.8\textwidth]{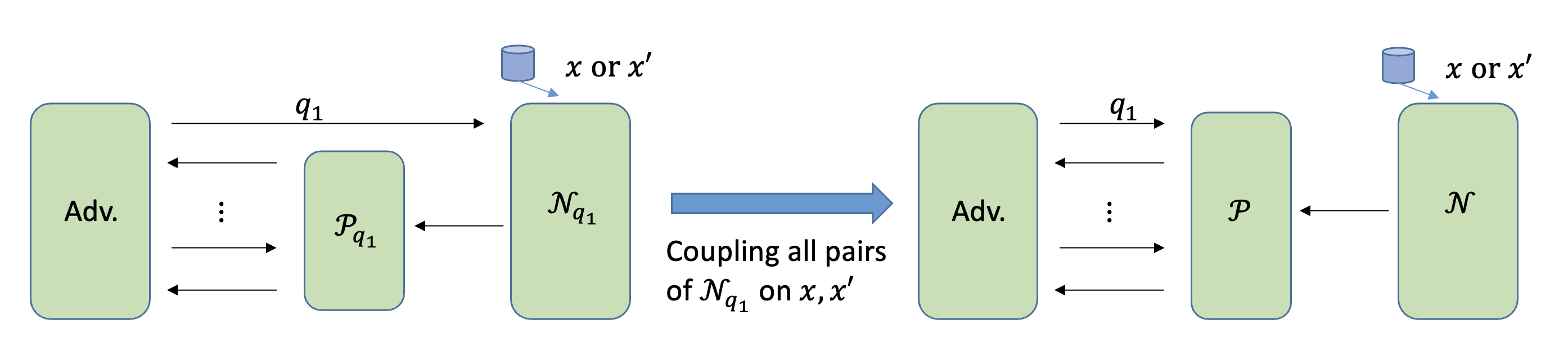}
	\caption{Case 1: The adversary $B$ sends the first query $q_1$.}\label{fig:coupling}
\end{figure}


\textbf{Case 2.} The mechanism $\mathcal{M}$ sends the first message $a_1$ to the adversary $B$. Fix a pair of neighboring datasets $x, x'$. We denote the mechanism conditioned on every $a_1$ by $\mathcal{M}_{a_1}$, and let $f_{a_1}$ be the trade-off function of  $\mathcal{M}_{a_1}$ (maximized over all adversaries). We denote the random variable of the first message as $A_1, A'_1$ on datasets $x,x'$, respectively. 
By induction, $\mathcal{M}_{a_1}$ can be simulated by a post-processing of a non-interactive $ f_{a_1}$-DP mechanism $\mathcal{N}_{a_1}$.
Thus, $\mathcal{M}$ can be simulated by a post-processing of the non-interactive mechanism $\mathcal{N}$ where $\mathcal{N}(x) \equiv (A_1, \mathcal{N}_{A_1}(x))$ and $\mathcal{N}(x') \equiv (A'_1, \mathcal{N}_{A'_1}(x'))$. We use the chain rule to argue that $T( \mathcal{N}(x), \mathcal{N}(x')  )\ge  \mathrm{ChainRule}_{A_1,A'_1}\left(  (f_{a_1})_{a_1\in \supp(A_1)\cap\supp(A'_1)} \right) =f$.
We conclude that Theorem \ref{thm.reduction} holds for $k$ rounds of interactions. 

\begin{figure}[H]
\centering
\includegraphics[width=0.8\textwidth]{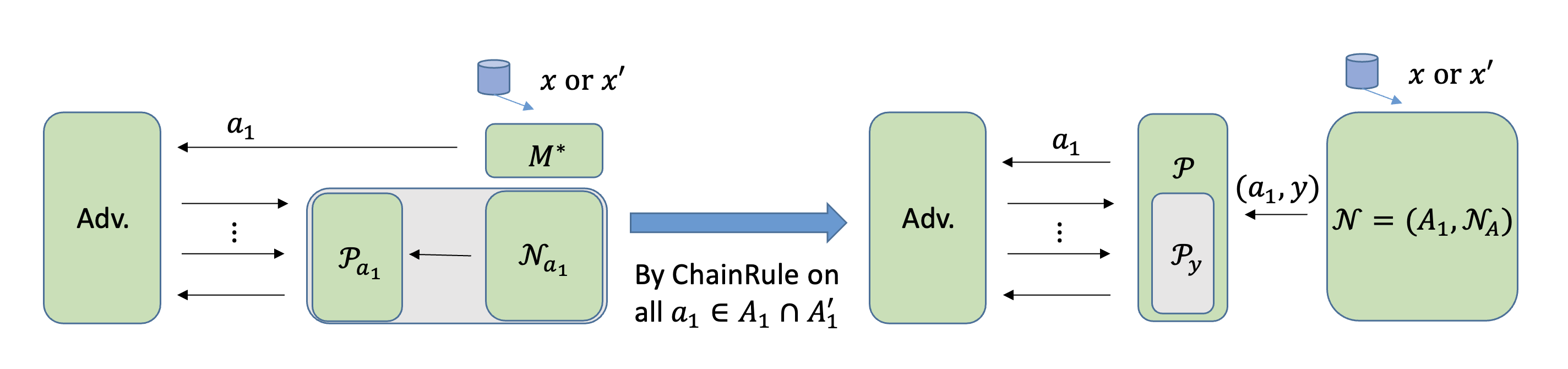}
\caption{Case 2: The mechanism $\mathcal{M}$ sends the first message $a_1$. }\label{fig:chain}
\end{figure}

\subsection{Independent Work by Lyu}

In independent and concurrent work, Lyu~\cite{lyu2022composition} proves Theorem \ref{thm.concomp} with a different argument.  They show
that every interactive $(\eps,\delta)$-DP mechanism can be simulated by interactive post-processing of a non-interactive $(\eps,\delta)$-DP mechanism, via an argument that is specific to $(\eps,\delta)$-DP that does not seem to generalize to arbitrary tradeoff functions $f$.  Indeed, they leave the the general case of $f$-DP as an open problem, which is solved by our Theorems~\ref{thm.fdp} and \ref{thm.reduction}.

On the other hand, Lyu~\cite{lyu2022composition} also proves an optimal concurrent composition theorem for R\'enyi DP of any fixed order. In an earlier version of our paper~\cite{vadhan2022concurrent}, we also claimed such a result, but our proof was incorrect (except for the case of R\'enyi DP of order $\alpha=1$),\footnote{Specifically, we stated and used a chain rule for R\'enyi divergence that only holds for order $\alpha=1$ (i.e. KL divergence).} as pointed out to us by Lyu. In this revision, we give a simple proof of Lyu's theorem for R\'enyi DP by characterizing the optimal adversary strategy in Section \ref{sec.RDP}.


\section{Generalized Definitions of DP Mechanisms}\label{sec.dDP}


To prove our results and discuss the several variants of differential privacy, it is convenient to introduce a more general abstraction, where distances between probability distributions can be in an arbitrary partially ordered set.


\begin{definition}[Generalized probability distance]
	A {\em generalized probability distance measure} is a tuple $(\mathcal{D}, \preceq, D)$ such that
	\begin{enumerate}
		\item $(\mathcal{D}, \preceq)$ is a partially ordered set (poset).
		\item $D$ is a mapping that takes any two random variables $X,X'$ over the same measurable space to an element $D(X,X')$ of $\mathcal{D}$.
		\item (Post-processing.) The generalized distance mapping $D$ is closed under post-processing, meaning that for every function $g$, $D(g(X), g(X'))\preceq D(X,X')$.
		\item (Joint Convexity.)  Suppose we have a collection of random variables $(X_i, X'_i)_{i\in \mathcal{I}}$ and a random variable $I$ distributed on $\mathcal{I}$. If $D(X_i, X'_i) \preceq d$ for all $i\in \mathcal{I}$, then $D(X_I, X'_I) \preceq d$.
	\end{enumerate}
\end{definition}

For the generalized notion $d$-$\mathcal{D}$ DP, the difficulty of distinguishing two neighboring datasets is measured by the generalized distance between the distributions of an adversary's views. The partially ordered set allows us to compare the level of privacy guarantees of mechanisms.


\begin{definition}[$d$-$\mathcal{D}$ DP]\label{def.ddp}
	Let $(\mathcal{D}, \preceq, D)$ be a generalized probability distance.
	For $d \in \mathcal{D}$, we call an interactive mechanism $\mathcal{M}$ {\em $d$-$\mathcal{D}$ DP} if for every adversary $B\in \mathcal{B}$ and every pair of neighboring datasets $x,x'$, we have 
	$$D(\View ( B \leftrightarrow \mathcal{M}(x))   ,\View (B \leftrightarrow \mathcal{M}(x') )) \preceq d.$$ 
\end{definition} 


Let us instantiate the standard pure DP and its variants using the definition above by specifying the generalized distances.

\paragraph{Example: pure DP.} For pure DP,  a smaller $\eps$ provides stronger privacy guarantee, so the partially ordered set  $\mathcal{D}$ is defined as $((\mathbb{R}^{\ge0})\cup \{\infty\}, \le)$. 
The distance mapping is the max-divergence $D_\infty$.  For two probability distributions $P$ and $Q$, the max-divergence is $$D_\infty(P||Q):=\sup_{T\subset \supp(P)}\log \left(  \frac{\Pr(P(x)\in T)}{\Pr(Q(x)\in T)} \right) .$$
Max-divergence is closed under post-processing due to the data-processing inequality. Max-divergence satisfies joint convexity due to the following lemma.
\begin{lemma}[\cite{van2014renyi}]\label{lemma.convex}
	For every two pairs of probability distributions $(P_0, Q_0)$ and $(P_1, Q_1)$, and every $\lambda\in (0,1)$, 
	\begin{equation*}
		D_\infty((1-\lambda) P_0+\lambda P_1 || (1-\lambda) Q_0+Q_1)\le \max \{ D_\infty(P_0||Q_0) , D_\infty(P_1||Q_1) \}.
	\end{equation*}
\end{lemma}

\paragraph{Example: R\'enyi DP} For R\'enyi DP of order $\alpha$, the partially ordered set $\mathcal{D}$ is also $((\mathbb{R}^{\ge0})\cup \{\infty\}, \le)$. The distance mapping is $\alpha$-R\'enyi divergence for $\alpha\in(1, \infty)$. 
The R\'enyi divergence is defined as follows.
\begin{definition}[R\'enyi divergence \cite{renyi1961measures}]
	For two probability distribution $P$ and $Q$, the R\'enyi divergence of order $\alpha>1$ is 
	\begin{equation*}
		D_\alpha(P||Q)=\frac{1}{\alpha-1}\log \left(\mathrm{E}_{x\sim Q} \left[\left(\frac{P(x)}{Q(x)}\right)^\alpha\right]\right).
	\end{equation*}
\end{definition}
R\'enyi divergence is also closed under post-processing due to the data-processing inequality, and it satisfies the joint convexity because an analogue of Lemma \ref{lemma.convex} also holds for R\'enyi divergence:

\begin{lemma}[\cite{van2014renyi}]
	For every order $\alpha>1$, every two pairs of probability distributions $(P_0, Q_0)$ and $P_1, Q_1$, and every $\lambda\in (0,1)$, 
	\begin{equation*}
		D_\alpha((1-\lambda) P_0+\lambda P_1 || (1-\lambda) Q_0+Q_1)\le \max \{ D_\alpha(P_0||Q_0) , D_\infty(P_1||Q_1) \}.
	\end{equation*}
\end{lemma}

\paragraph{Example: $f$-DP.} For $f$-DP, the partially ordered set  $\mathcal{D}$ is defined as $(\mathcal{F}, \preceq)$, where $\mathcal{F}$ is the set of all trade-off functions that satisfies the conditions in Proposition \ref{def.tradeoffclass}. The partial ordering is defined as $f_1\preceq f_2$ if  $f_1(\alpha)\ge f_2(\alpha)$ holds  for all $\alpha \in [0,1]$. Note that the direction of the inequalities is reversed, corresponding to the fact that a larger trade-off function means less privacy loss. The distance mapping is the trade-off function $T$ in Definition \ref{tradeoff.def}.

$f$-DP also satisfies the two properties. First, $f$-DP is preserved under post-processing. We will only need to show the joint convexity of $f$-DP. 

\begin{lemma}
	Suppose we have a collection of random variables $(X_i, X'_i)_{i\in \mathcal{I}}$ and a random variable $I$ distributed on $\mathcal{I}$. If $T(X_i, X'_i)\ge f$ for all $i\in \mathcal{I}$, then $T(X_I, X'_I)\ge f$.
\end{lemma}

\begin{proof}
	For any random variable $I$ distributed on $\mathcal{I}$.  We have
	\begin{align}
		\MoveEqLeft	T(X_I, X'_I)(\alpha)=\inf_{\phi} \left\lbrace  \mathrm{E}\left[ 1-\phi(X'_I)  \right] :  \mathrm{E}[\phi(X'_I)]\le \alpha   \right\rbrace   \notag \\
		&=\inf_{\phi}\left\lbrace \mathrm{E}_{i\sim I} \mathrm{E} [1-\phi(X'_i)]:  \mathrm{E}_{i\sim I} \mathrm{E}[\phi(X_i)]\le \alpha  \right\rbrace  \notag  \\
		&\ge \inf_{\phi} \left\lbrace   \mathrm{E}_{i\sim I} [f(\mathrm{E}[\phi(X_i)]  )]:     \mathrm{E}_{i\sim I} \mathrm{E}[\phi(X_i)]\le \alpha      \right\rbrace  \tag{$f$ non-decreasing}  \\
		&\ge \inf_{\phi} \left\lbrace f(\mathrm{E}_{i\sim I} \mathrm{E}[\phi(X_i)]  ): \mathrm{E}_{i\sim I} \mathrm{E}[\phi(X_i)]\le \alpha      \right\rbrace  \tag{$f$ convex}  \\
		&= f(\alpha). \notag
	\end{align}
	Therefore, we have $T(X_I, X'_I)\ge f$. 
\end{proof}

It is useful to work with distance posets that are {\em complete}:
\begin{definition}[Complete poset]
	A partially ordered set (poset) $(\mathcal{D}, \preceq)$ is {\em complete} if for every nonempty subset $S \subseteq \mathcal{D}$ has a supremum $\sup(S)$, where $s \preceq \sup(S)$ for every $s \in S$, and $\sup(S)\preceq t$ for every $t$ satisfying $s \preceq t$ for every $s \in S$.  
\end{definition}

We note that $\sup(S)$ is always unique. The poset $((\mathbb{R}^{\ge0})\cup \{\infty\}, \le)$ used in pure DP and R\`enyi DP is complete by the usual completeness of the real numbers. For the poset $(\mathcal{F}, \preceq)$ used in $f$-DP, we prove it below. Note that if $(\mathcal{D}, \preceq)$ is complete then in Definition \ref{def.ddp} we can take $d=\sup_B D(\View ( B \leftrightarrow \mathcal{M}(x))   ,\View (B \leftrightarrow \mathcal{M}(x') ))$ as the optimal privacy loss for a given interactive mechanism $\mathcal{M}$.


\begin{lemma}\label{lemma.complete}
	The partially ordered set $(\mathcal{F}, \preceq)$, where $\mathcal{F}$ consists of all trade-off functions satisfying the conditions in Proposition \ref{def.tradeoffclass}, is complete. Specifically, for $S \subseteq \mathcal{F}$, $\sup S$ is a trade-off function defined as follows.
	\begin{equation}
	\sup S = h(\alpha) :=\inf_{\substack{F: \supp(F)\subseteq S \\ A: S\rightarrow [0,1]}} \left\lbrace \mathrm{E}[F(A(F))]: \mathrm{E}[A(F)]\le \alpha \right\rbrace,
	\end{equation}
	where $F$ is a random variable that takes value in $S$ and $A: S\rightarrow [0,1]$ is a function. 
\end{lemma}


%

\begin{proof}
	We first show that $h$ is the least upper bound for $S$. We shall show that for any tradeoff function $h'$ such that $f \preceq h' $ for every $f\in S$, we have $h \preceq h' $. Let $F$ be a random variable such that $\supp(F)\subseteq S$, and let $A: S\rightarrow [0,1]$ be a function such that $ \mathrm{E}[A(F)] \le \alpha$. As stated in Proposition \ref{def.tradeoffclass}, a trade-off function is convex and non-increasing, so by Jensen's inequality, we have 
	\begin{equation*}
		h'(\alpha)\le h'(\mathrm{E}[A(F)]) \le \mathrm{E}[h'(A(F))].
	\end{equation*}
	By the definition of the partial ordering, we have $h'(\alpha)\le f(\alpha)$ for every $\alpha \in [0,1]$ and every $f\in S$, so $ \mathrm{E}[h'(A(F))]\le  \mathrm{E}[F(A(F))]$. Therefore, we have 
	\begin{equation*}
		h'(\alpha)\le \mathrm{E}[F(A(F))],
	\end{equation*}
	Taking the infimum over $F$ and $A$ on both sides, we get $h'(\alpha)\le h (\alpha)$, and therefore, $h\preceq h'$.
	
	Next, we shall show that $h$ is a trade-off function. Following the proposition \ref{def.tradeoffclass}, it suffices to check the four properties for $h$. We begin with proving the convexity of $h$.
	For every $a, b\in[0,1]$, and every $\lambda\in[0,1]$, we have 
	\begin{align}
		h (\lambda a + (1-\lambda) b)=&\inf_{F, A} \left\lbrace  \mathrm{E}[F(A(F))]: \mathrm{E}[A(F)]\le \lambda a + (1-\lambda) b \right\rbrace \notag \\
		\le &\lambda \inf_{F, A} \left\lbrace  \mathrm{E}[F(A(F))]: \mathrm{E}[A(F)]\le a \right\rbrace  
		+(1-\lambda)\inf_{F, A} \left\lbrace  \mathrm{E}[F(A(F))]: \mathrm{E}[A(F)]\le  b \right\rbrace    \label{ineq.1}\\
		=& \lambda h(a) + (1-\lambda) h(b).\notag
	\end{align}
	where inequality \eqref{ineq.1} is because that for every $A_a,F_a$ that satisfies $\mathrm{E}[A(F)]\le a$ and every $A_b,F_b$ that satisfies $\mathrm{E}[A(F)]\le b$, the linear combination $A(F)=\lambda A_a(F_a)+(1-\lambda)A_b(F_b)$ satisfies $\mathrm{E}[A(F)]\le \lambda a + (1-\lambda) b $.  
	Thus, $h$ is convex. $h$ is non-increasing and continuous on $[0,1]$ due to the monotonicity and continuity of $f\in \mathcal{F}$ (Proposition \ref{def.tradeoffclass}). Finally, since $f(x)\le 1-x$ for every $f\in \mathcal{F}$, we have 
	$$h(\alpha)\le \mathrm{E}[F(A(F))] \le \mathrm{E}[1-A(F)] \le 1-\alpha.$$
	Therefore, $h$ is a trade-off function, and $\sup S$ exists.
	
\end{proof}

A convenient consequence of joint convexity is that it suffices to consider deterministic adversaries.

\begin{lemma}\label{lemma.det}
	An interactive mechanism $\mathcal{M}$ is $d$-$\mathcal{D}$ DP, if and only if for every pair of neighboring datasets $x,x'$, for every deterministic adversary algorithm $B$, we have $D(\View ( B, \mathcal{M}(x))   ,\View (B, \mathcal{M}(x') )) \preceq d$. 
\end{lemma}

\begin{proof}
	The necessity is immediately implied by Definition \ref{def.ddp}. We shall prove the sufficiency. Let $B$ be a randomized adversary.  If we fix the coin tosses of
	$B$ to a value $r$, we obtain a deterministic adversary $B_r$.
	By hypothesis, we have $D(\View ( B_r, \mathcal{M}(x))   ,\View (B_r, \mathcal{M}(x') )) \preceq d$.  Now let random variable $R$ be uniformly distributed over the coins of $A$.  Then the view of the randomized adversary $B$ when interacting with $\mathcal{M}$ consists of the coins $R$ and the view of the deterministic adversary $B_R$.  That is,
	$$\View ( B \leftrightarrow \mathcal{M}(x)) = (R, \View(B_R\leftrightarrow \mathcal{M}(x))),$$
	and similarly for $x'$. 
	By joint convexity, we deduce:
	$$\View ( B \leftrightarrow \mathcal{M}(x)) \preceq d.$$
\end{proof}



\section{Coupling and Chain Rule Properties of $f$-DP}

In this section, we prove that $f$-DP has the coupling and chain rule properties that we use to prove Theorems \ref{thm.fdp} and \ref{thm.reduction}.


\begin{definition}[Coupling property]\label{def.couplingP}
	We say that a generalized distance $D$ has the {\em coupling property} if for any two pairs of random variable $X,X'$ and $Y,Y'$,  we have $D(X, X') \preceq d$ and $D(Y, Y') \preceq d$, then there exists a coupling of $X$ and $Y$ (denoted as $(X,Y)$), and a coupling of $X'$ and $Y'$ (denoted as $(X',Y')$), such that $D((X,Y),(X',Y')) \preceq d$.
\end{definition}


\begin{lemma}[Lemma \ref{lemma.coupling} restated]\label{lemma.coupling2}
	$(\mathcal{F}, \preceq, T)$ has the coupling property: Suppose $f$ is a trade-off function and
	we have random variables $X$, $Y$  and $X'$, $Y'$ such that 
	\begin{equation*}
		T(X,X') \ge f \quad \text{and} \quad T(Y,Y')\ge f.
	\end{equation*}
	Then there exists couplings $(X,Y)$ and $(X',Y')$ such that
	\begin{equation*}
		T((X,Y)||(X',Y'))\ge f.
	\end{equation*}
\end{lemma}

We prove this lemma using the following result:

\begin{theorem}[Blackwell Theorem \cite{dong2019gaussian} (also see \cite{blackwell1953equivalent, kairouz2015composition})]
	Let $P, Q$ be probability distributions on $X$ and $P', Q'$ be probability distributions on $Y$. The following two statements are equivalent:
	\begin{enumerate}
		\item $T(P,Q) \le T(P',Q')$.
		\item There exists a randomized algorithm $\mathrm{Proc}: X\rightarrow Y$ such that $\mathrm{Proc}(P)=P'$ and  $\mathrm{Proc}(Q)=Q'$.
	\end{enumerate}
\end{theorem}

\begin{proof}[Proof of Lemma \ref{lemma.coupling2}]
	
	Since a function is called a trade-off function if it is equal to $T(P,P')$ for some distribution $P$ and $P'$, for a given rade-off function $f$, there exists a pair of random variables $P, P'$  such that $T(P, P')=f$. 
	By the Blackwell Theorem, since $T(P, P')=f\le T(X, X')$, there exists a randomized algorithm $\mathcal{P}_0$ such that $\mathcal{P}_0(P)$ and  $\mathcal{P}_0(P')$ are identically distributed to $X$ and  $X'$, respectively. Similarly, since $T(P, P')=f\le T(Y, Y')$, there exists a randomized algorithm $\mathcal{P}_1$ such that $\mathcal{P}_1(P), \mathcal{P}_1(P')$ is identically distributed to $Y, Y'$, respectively. We construct a coupling of $X$ and $Y$ as $(\mathcal{P}_0(P), \mathcal{P}_1(P))$, and  a coupling of $X'$ and $Y'$ as $(\mathcal{P}_0(P'), \mathcal{P}_1(P'))$. Then the trade-off function between the two couplings satisfies the following inequality.
	\begin{align}
		&T((\mathcal{P}_0(P), \mathcal{P}_1(P)), (\mathcal{P}_0(P'), \mathcal{P}_1(P')))  \notag \\
		&\ge T(P,P') \label{eq2} \\
		&=f,
	\end{align}
	where Equation \eqref{eq2} follows from Lemma 2.9 in \cite{dong2019gaussian}, completing the proof.
\end{proof}

To formally state the chain rule property, we need a couple of definitions.



\begin{definition}[Continuous function]\label{def.conf}
	Let $(A, \preceq)$ and $(B, \preceq)$ be complete posets. A function $f: A\rightarrow B$ is {\em continuous} if $f(\sup(S))=\sup(f(S))$ for every set $S\subseteq A$.
\end{definition}

Observe that every continuous function is monotone: if $a\preceq a'$ are elements of $A$, then $f(a')=f(\sup(a, a'))=\sup(f(a),f(a'))\succeq f(a)$.

\begin{definition}[Continuous in each variable]\label{def.conf2}
	Let $S$ be a finite set and $|S|=n$. Let $(A, \preceq)$ and $(B, \preceq)$ be complete posets. A function $f: A^S\rightarrow B$ is {\em continuous in each variable} if for every $i$, 
	and
	for every $a_1,\ldots, a_{i-1}, a_{i+1}, \ldots, a_n \in A$, 
	the function $g(x)=f(a_1,\ldots,a_{i-1},x,a_{i+1},\ldots,a_n)$ is a continuous function from $A$ to $B$.   
\end{definition}

\begin{definition}[Chain rule]\label{def.chain}
	We say that a generalized probability distance $(\mathcal{D}, \preceq, D)$ satisfies the chain rule property if
	for every pair of random variables $(X, X')$ on the same domain $\mathcal{X}$, there is a function that is continuous in each variable:
	$ \mathrm{ChainRule}_{X,X'}: \mathcal{D}^{\supp(X)\cap \supp(X')}\rightarrow \mathcal{D}$ such that for every pair of random variables $Y$ and $Y'$ where $Y$ is jointly distributed with $X$ and $Y'$ is jointly distributed with $X'$, we have
	$$D((X,Y),(X',Y')) = \mathrm{ChainRule}_{X,X'}((D(Y|X=x,Y'|X'=x)_{x \in \supp(X) \cap \supp(X')}).$$
\end{definition}

As an example, the standard chain rule of KL divergence is as follows.
\begin{align*}
	D_{\mathrm{KL}}((X,Y)||(X',Y'))&=D_{\mathrm{KL}}(X||X')+D_{\mathrm{KL}}(Y|X|| Y'|X')\\
	&=D_{\mathrm{KL}}(X||X')+\mathrm{E}_{x\sim X}D_{\mathrm{KL}}(Y|X=x|| Y'|X'=x).
\end{align*}
So fixing $X$ and $X'$, we can calculate the KL divergence for arbitrary $Y$ and $Y'$ as a function of the KL divergences $\mathrm{KL}(Y|X=x|| Y'|X'=x)$.

In Lemma \ref{lemma.chain2}, we show that $f$-DP has the chain rule property.

\begin{lemma}[Lemma \ref{lemma.chain} restated] \label{lemma.chain2}
	For every pair of random variables $X, X'$ with finite support, there exists a function that is continuous in each variable $\mathrm{ChainRule}_{X,X'}$ such that for every random variable $Y$ jointly distributed with $X$, and every random variable $Y'$ jointly distributed with $X'$, we have 
	\begin{equation*}
		T((X,Y), (X',Y'))=\mathrm{ChainRule}_{X,X'}((T(Y|X=x, Y'|X'=x)))_{x\in \supp(X)\cap \supp(X') }).
	\end{equation*}
	where $T$ is a trade-off function.
\end{lemma}



\begin{proof}
	
	\begin{claim}
		The ChainRule function for $f$-DP is given as follows.
		\begin{equation}\label{eq.f_chain}
			\mathrm{ChainRule}_{X,X'}\left( (f_x)_{x\in \supp(X)\cap \supp(X') }\right)(\alpha) =\inf_{\alpha_x\in[0,1]}\left\lbrace  \mathrm{E}_{x\sim X'}[f_x(\alpha_x)]: \mathrm{E}_{x\sim X}[\alpha_x]\le \alpha  \right\rbrace .
		\end{equation}
	\end{claim} 
	
	We first prove this claim.	
	Suppose $Y$ is jointly distributed with $X$, and $Y'$ is jointly distributed with $X'$.  We consider hypothesis tests distinguishing $(X,Y)$ and $(X',Y')$. Let $\phi$ be any decision rule for this testing, $\alpha(\phi)$ and $\beta(\phi)$ be the corresponding Type I error and Type II error, respectively. For a given instance $x \in \supp(X)\cap \supp(X')$, let $\phi_x(y):=\phi(x,y)$. Additionally, let $f_x$ be the trade-off function conditioned on $x$, i.e.,
	
	\begin{equation}\label{def.fx0}
		f_{x}(\alpha):=T(Y|X=x, Y'|X'=x)(\alpha)  . 
	\end{equation}
	The type I error $\alpha_\phi$ and type II error  $\beta_\phi$ are given as 
	\begin{equation*}
		\alpha_\phi=\mathrm{E}[\phi(x,y)]=\mathrm{E}_{x\sim X}\mathrm{E}_{y\sim Y}[\phi_x(y)],
	\end{equation*}
	and 
	\begin{equation*}
		\beta_\phi=1- \mathrm{E}[\phi(x',y')]=1-\mathrm{E}_{x\sim X'} \mathrm{E}_{y'\sim Y'}[\phi_{x'}(y')].
	\end{equation*}	
	For every fixed $x \in \supp(X)\cap \supp(X')$ and every decision rule $\phi$ such that $\mathrm{E}_{y\sim Y}[\phi_x(y)]=\alpha_x$, by the definition of $f_{x}$ in \eqref{def.fx0}, we have 
	\begin{equation*}
		1- \mathrm{E}_{y'\sim Y'}[\phi_{x'}(y')] \ge f_{x}(\alpha_x).
	\end{equation*}
	Therefore, the trade-off function between $(X,Y)$ and $(X',Y')$ satisfies the following inequality:
	\begin{align}
		T((X,Y),(X',Y'))(\alpha)=\inf_\phi \left\lbrace  \beta_\phi:\alpha_\phi\le \alpha  \right\rbrace  &= \inf_\phi \left\lbrace  1-\mathrm{E}_{x\sim X'} \mathrm{E}_{y'\sim Y'}[\phi_{x'}(y')]: \mathrm{E}_{x\sim X}\mathrm{E}_{y\sim Y}[\phi_x(y)]\le \alpha \right\rbrace  \notag \\ 
		&\ge \inf_{\{\alpha_x\}}\left\lbrace \mathrm{E}_{x\sim X'}[f_{x}(\alpha_x)]: \mathrm{E}_{x\sim X}[\alpha_x]\le \alpha \right\rbrace . \label{eq.5}
	\end{align}
	On the other hand, by the definition of $f_{x}$, for every $0<\alpha_x<1$ and $\delta>0$, there exists a decision rule $\phi^\delta$ such that $1- \mathrm{E}_{y'\sim Y'}[\phi^\delta_{x'}(y')]  \le f_{x}(\alpha_{x})+\delta$ and $\mathrm{E}_{y\sim Y}[\phi^\delta_x(y)]\le \alpha_x$. Then we have
	\begin{align*}
		\inf_{\{\alpha_x\}} \left\lbrace \mathrm{E}_{x\sim X'}[f_{x}(\alpha_{x})]: \mathrm{E}_{x\sim X}[\alpha_{x}]\le \alpha \right\rbrace &\ge \inf_\delta \left\lbrace \mathrm{E}_{x\sim X'}[1- \mathrm{E}_{y'\sim Y'}[\phi^\delta_{x'}(y')]-\delta]: \mathrm{E}_{x\sim X}[ \mathrm{E}_{y\sim Y}[\phi^\delta_{x}(y)]]\le \alpha \right\rbrace . \\
		&=\inf_\delta \left\lbrace  \beta_{\phi^\delta}:\alpha_{\phi^\delta}\le \alpha \right\rbrace -\delta \\
		&\ge \inf_\phi \left\lbrace  \beta_{\phi}:\alpha_{\phi}\le \alpha \right\rbrace -\delta \\
		&=T((X,Y),(X',Y'))(\alpha)-\delta.
	\end{align*}
	Let $\delta$ go to $0$, and combining with
	Equation \eqref{eq.5}, we have
	\begin{equation}\label{eq.7}
		T((X,Y),(X',Y'))(\alpha)=\inf \left\lbrace \mathrm{E}_{x\sim X'}[f_{x}(\alpha_{x})]: \mathrm{E}_{x\sim X}[\alpha_{x}]\le \alpha \right\rbrace .
	\end{equation}
	completing the proof for this claim.
	
	Next, we shall show that the ChainRule function defined in \eqref{eq.f_chain} is continuous in each variable. 
	Our goal is to show that for every $i$, every $S_i \subseteq \mathcal{D}$, and for every $f_1,\ldots, f_{i-1}, f_{i+1}, \ldots, f_n \in \mathcal{D}$, we have
	\begin{equation*}
		\mathrm{ChainRule}_{X,X'}(f_1, \ldots, f_{i-1}, \sup S_i, f_{i+1}, \ldots, f_n )= \sup(\mathrm{ChainRule}_{X,X'} (f_1,\ldots, f_{i-1}, f_i, f_{i+1}, \ldots, f_n): f_i\in S_i).
	\end{equation*}

	Let $S=(\mathrm{ChainRule}_{X,X'} (f_1,\ldots, f_{i-1}, f_i, f_{i+1}, \ldots, f_n): f_i\in S_i)$. For every random variable $F$ such that $\supp(F)\subseteq S$, we have $F$ taking values as $\mathrm{ChainRule}_{X,X'} (f_1,\ldots, f_{i-1}, f_i, f_{i+1}, \ldots, f_n): f_i\in S_i$, so it is equivalent to consider a random variable $F_i$ such that $\supp(F_i)\subseteq S_i$. For every function $A: S\rightarrow [0,1]$, we also slightly abuse the notation and use $A(F_i)$ to represent $A(F)$.
	For every $\alpha\in [0,1]$, we have 

	\begin{align}
	\MoveEqLeft	 \sup(\mathrm{ChainRule}_{X,X'} (f_1,\ldots, f_{i-1}, f_i, f_{i+1}, \ldots, f_n): f_i\in S_i)(\alpha) \notag \\
	=&\inf_{\substack{F_i: \supp(F_i)\subseteq S_i \\ A: S\rightarrow [0,1]}} \left\lbrace \mathrm{E}_{F_i}\left[ \mathrm{ChainRule}_{X,X'} (f_1,\ldots, f_{i-1}, F_i, f_{i+1}, \ldots, f_n) (A(F_i)) \right] : \mathrm{E}_{F_i}[A(F_i)]\le \alpha \right\rbrace \tag{by Lemma \ref{lemma.complete}} \\
	=&\inf_{\substack{F_i: \supp(F_i)\subseteq S_i \\ A: S\rightarrow [0,1]}} \left\lbrace \mathrm{E}_{F_i}\left[ \inf_{\substack{A_1,\ldots,A_n \\ A_j: S_i\rightarrow [0,1]  }} \left\lbrace \mathrm{E}_{j\sim X'} \left[ f_{j} (A_j(F_i)) \cdot \mathbb{I}(j\neq i) + F_i( A_i(F_i))\cdot \mathbb{I}(j=i)  \right] :  \mathrm{E}_{j\sim X}[A_j(F_i)]\le A(F_i)   \right\rbrace  \right] :  \right.  \notag \\
  &\left.	 \mathrm{E}_{F_i}[A(F_i)]\le \alpha \right\rbrace \tag{by \eqref{eq.f_chain}} \\
	=&\inf_{\substack{F_i: \supp(F_i)\subseteq S_i \\ A: S_i\rightarrow [0,1] \\ A_j: S_i\rightarrow [0,1] }} \left\lbrace \mathrm{E}_{F_i} \mathrm{E}_{j\sim X'}  \left[f_{j} (A_j(F_i)) \cdot \mathbb{I}(j\neq i) + F_i( A_i(F_i))\cdot \mathbb{I}(j=i)   \right] : \mathrm{E}_{j\sim X}[A_j(F_i)]\le A(F_i),   \mathrm{E}_{F_i}[A(F_i)] \le \alpha \right\rbrace. \label{eq.constraints1}
\end{align}	

We can interchange the expectation and the infimum in Equation~\eqref{eq.constraints1} because $A_j(f_i)$, $j=1, \ldots, n$, are independent across $f_i\in F_i$. We also have that
\begin{align}
	\MoveEqLeft	 \mathrm{ChainRule}_{X,X'}(f_1, \ldots, f_{i-1}, \sup S_i, f_{i+1}, \ldots, f_n )(\alpha) \notag \\
	=&\inf_{\alpha_1,\ldots, \alpha_n}\left\lbrace \mathrm{E}_{j\sim X'}\left[ f_{j} (\alpha_j) \cdot \mathbb{I}(j\neq i) + (\sup S_i)( \alpha_i)\cdot \mathbb{I}(j=i)  \right] :  \mathrm{E}_{j\sim X}[\alpha_j]\le \alpha   \right\rbrace \tag{by \eqref{eq.f_chain}}  \\
	=&\inf_{\alpha_1,\ldots, \alpha_n}\left\lbrace \mathrm{E}_{j\sim X'}\left[ f_{j} (\alpha_j) \cdot \mathbb{I}(j\neq i) + \inf_{\substack{F_i: \supp(F_i)\subseteq S_i \\ A_{i}: S_i\rightarrow [0,1]}}\left\lbrace \mathrm{E}_{F_i}\left[ F_i(A_{i}(F_i))  \right] : \mathrm{E}[A_{i}(F_i)]\le \alpha_i \right\rbrace \cdot \mathbb{I}(j=i)  \right] :  \mathrm{E}_{j\sim X}[\alpha_j]\le \alpha   \right\rbrace \tag{by Lemma \ref{lemma.complete}}  \\
    =&\inf_{\substack{F_i: \supp(F_i)\subseteq S_i \\ A_i: S_i\rightarrow [0,1] \\ \alpha_1, \ldots, \alpha_n }} \left\lbrace \mathrm{E}_{F_i} \mathrm{E}_{j\sim X'}  \left[f_{j} (\alpha_j) \cdot \mathbb{I}(j\neq i) + F_i( A_i(F_i))\cdot \mathbb{I}(j=i)   \right] :  \mathrm{E}_{j\sim X}[\alpha_j]\le \alpha,  \mathrm{E}_{F_i}[A_i(F_i)] \le \alpha_i   \right\rbrace. \label{eq.constraints2}
\end{align}

The constraints in \eqref{eq.constraints1} are equivalent to the constraints in \eqref{eq.constraints2}, which can be seen as follows. For all $\alpha_j(F_i)$ ($j=1,\ldots, k$, and $F_i\in \supp(S_i)$) satisfy the constraints in \eqref{eq.constraints1}, let $\alpha_j=\mathrm{E}_{F_i}[ A_j(F_i) ]$, then $\alpha_j$ satisfy the constraints in \eqref{eq.constraints2}. On the other hand, for all $\alpha_j$ ($j=1,\ldots, k$) satisfy the constraints in \eqref{eq.constraints2}, let $\alpha_j=\mathrm{E}_{F_i}[ A_j(F_i) ]$, then $\alpha_j(F_i)$ satisfy the constraints in \eqref{eq.constraints1}. Moreover, with $\alpha_j=\mathrm{E}_{F_i}[ A_j(F_i) ]$, we have
\begin{align}
\MoveEqLeft \mathrm{E}_{j\sim X'}  \left[f_{j} (\alpha_j) \cdot \mathbb{I}(j\neq i) \right]  \notag  \\
=&  \mathrm{E}_{j\sim X'}  \left[f_{j} ( \mathrm{E}_{F_i}[ A_j(F_i) ] ) \cdot \mathbb{I}(j\neq i) \right] \notag \\
\le& \mathrm{E}_{j\sim X'}  \mathrm{E}_{F_i}  \left[f_{j} (  A_j(F_i) ) \cdot \mathbb{I}(j\neq i) \right], \label{eq.convex}
\end{align}
where \eqref{eq.convex} is because trade-off functions are convex. Hence, 
\begin{equation*}
\inf_{\substack{F_i: \supp(F_i)\subseteq S_i \\ A_i: S_i\rightarrow [0,1] \\ \alpha_1, \ldots, \alpha_n }} \mathrm{E}_{j\sim X'}  \left[f_{j} (\alpha_j) \cdot \mathbb{I}(j\neq i) \right]  \le \inf_{\substack{F_i: \supp(F_i)\subseteq S_i \\ A: S_i\rightarrow [0,1] \\ A_j: S_i\rightarrow [0,1]  }} \mathrm{E}_{j\sim X'}  \mathrm{E}_{F_i}  \left[f_{j} (  A_j(F_i) ) \cdot \mathbb{I}(j\neq i) \right].
\end{equation*}
On the other hand, by setting $A_j(F_i)=\alpha_j$ for all $F_i\in S_i$, the above equal sign is reached.

Therefore, we have 
\begin{equation*}
	\mathrm{ChainRule}_{X,X'}(f_1, \ldots, f_{i-1}, \sup S_i, f_{i+1}, \ldots, f_n )= \sup(\mathrm{ChainRule}_{X,X'} (f_1,\ldots, f_{i-1}, f_i, f_{i+1}, \ldots, f_n): f_i\in S_i).
\end{equation*}	
	
\end{proof}

\section{Concurrent Composition of $d$-$\mathcal{D}$ DP}\label{sec.concompddp}


Theorem \ref{thm.post_gen} shows that if the generalized distance satisfies the coupling property in Definition \ref{def.couplingP}  and the chain rule in Definition \ref{def.chain} , then every interactive $d$-$\mathcal{D}$ DP mechanism can be simulated by an interactive post-processing of a non-interactive $d$-$\mathcal{D}$ DP mechanism. This is a generalized statement of Theorem \ref{thm.reduction}, as $f$-DP is an example of $d$-$\mathcal{D}$ DP.

\begin{theorem}[Theorem \ref{thm.reduction} generalized]\label{thm.post_gen}
	
	Assume that the generalized probability distance measure $(\mathcal{D}, \preceq, D)$ satisfies 
	\begin{enumerate}
		\item $(\mathcal{D}, \preceq)$ is complete.
		\item $\mathcal{D}$ satisfies the chain rule.
		\item every $d\in \mathcal{D}$ satisfies the coupling property.
	\end{enumerate}
	Then for every $d\in 
	\mathcal{D}$ and every interactive $d$-$\mathcal{D}$ DP mechanism $\mathcal{M}$ with finite communication complexity, and every pair of two neighboring datasets $x$ and $x'$, there exists a pair of random variables $Y, Y'$ and an randomized interactive post-processing mechanism $\mathcal{P}$ such that $D(Y, Y')\preceq d$, and for every adversary $B\in \mathcal{B}$, we have
	
	\begin{align}
		\View(B\leftrightarrow \mathcal{M}(x))&\equiv \View(B \leftrightarrow \mathcal{P}(Y))\\
		\View(B \leftrightarrow \mathcal{M}(x'))&\equiv \View(B \leftrightarrow \mathcal{P}(Y')).
	\end{align}
	
\end{theorem}


Note that the theorem is stated for mechanisms with {\em finite communication}, which is formally defined as follows.

\begin{definition}
	Let $(A, B)$ be an interactive protocol (as in Definition \ref{def.protocol}). 	We say that $A$ has {\em finite communication} if for every $x_A$ there is a constant $c$, such that for all $r_A, m_1, \ldots, m_{i-1}$, we have 
	\begin{enumerate}
		\item If $\max \{i, |m_1|, \ldots, |m_{i-1}|\}>c$, then $A(x_A, m_1, m_3, \ldots, m_{i-1}; r_A)=\texttt{halt}$.
		\item If $\max \{i, |m_1|, \ldots, |m_{i-1}|\}\le c$, then $ \sum_{j=0}^{i-1} \left| A(x_A, m_1, m_3, \ldots, m_j; r_A)\right| \le c$. 
	\end{enumerate}
	Here $|y|$ denotes the bit length of string $y$. $B$ having finite communication is defined symmetrically.
\end{definition}


\begin{proof}[Proof of Theorem \ref{thm.post_gen}]
	
	Our strategy is to apply the induction argument by the number of rounds of interactions. Fix a pair of neighboring datasets $x, x'$. We consider two cases depending on whether the first message sent from the mechanism $\mathcal{M}$ or the adversary $B$.

	\paragraph{Case 1.}  The adversary $B$ sends the first query $q_1$ to the mechanism $\mathcal{M}$. Fixing $q_1$, the subsequent interactive mechanism $\mathcal{M}_{q_1}$ with input $x$ is defined by
	$$\mathcal{M}_{q_1}(x, q_2, \ldots, q_m, r)=\mathcal{M}(x, q_1, \ldots, q_m, r).$$
	We claim that $\View ( A \leftrightarrow \mathcal{M}_{q_1} )$ consists of $m-1$ messages, and $\mathcal{M}_{q_1}$ satisfies $d$-$\mathcal{D}$ DP on the two neighboring datasets $x$ and $x'$. By induction, there exists a randomized interactive post-processing $\mathcal{P}_{q_1}$ and a pair of random variables $Y_{q_1}, Y'_{q_1}$ such that 
	$$D(Y_{q_1}, Y'_{q_1}) \preceq d,$$
	and 
	$$\mathcal{P}_{q_1}(Y_{q_1})\equiv \mathcal{M}_{q_1}(x) \quad \mathcal{P}_{q_1}(Y'_{q_1})\equiv \mathcal{M}_{q_1}(x').$$
	By coupling property, there exists a pair of random variables $Y$, $Y'$ and a randomized post-processing function $\mathcal{Q}_{q_1}$ such that $D(Y,Y')\preceq d$, and we have that
	$$\mathcal{Q}_{q_1}(Y)=Y_{q_1},$$
	$$\mathcal{Q}_{q_1}(Y')=Y'_{q_1}.$$
	So $Y$ and $Y'$ are produced by coupling all possible queries.
	Then the interactive post-processing $\mathcal{P}$ is defined by $\mathcal{P}_{q_1}\circ \mathcal{Q}_{q_1}$, i.e.,
	$$\mathcal{P}(y, q_1, q_2, \ldots, q_m)=\mathcal{P}_{q_1}(\mathcal{Q}_{q_1}(y), q_2, \ldots, q_m).$$


	\paragraph{Case 2.} The mechanism $\mathcal{M}$ sends the first message $a_1$ to the adversary $B$. Let $q_1,\ldots q_{m-1}$ be the queries from the adversary, and $A_1,\ldots,A_m$ be messages from the mechanism. Fixing $A_1=a_1$, the subsequent interactive mechanism $\mathcal{M}_{a_1}$ is defined by 
	$$\mathcal{M}_{a_1}(x, q_1, \ldots, q_{m-1}; g_x(r))=\mathcal{M}(x, q_1, \ldots, q_{m-1}; r).$$
	
	$\mathcal{M}_{a_1}$ uses its randomness to choose uniformly from randomness of $\mathcal{M}$ conditioned on $\mathcal{M}(x)=a_1$. Specifically, let $g_x$ be a random transformation such that if $R$ is uniform random for $\mathcal{M}$, then for all $x$, $g_x(R)$ is uniform on the randomness of $\mathcal{M}$ conditioned on $\mathcal{M}(x)=a_1$.
	
	We define the subsequent adversary $B_{a_1}(A_2, \ldots, A_m)=B(a_1, A_2, \ldots, A_m)$.
	We know that for all adversary strategy $B\in \mathcal{B}$, we have $D(\View ( B \leftrightarrow \mathcal{M}(x) ), \View ( B \leftrightarrow \mathcal{M}(x') )) \preceq d$, so we have $$\sup_B D(\View ( B\leftrightarrow \mathcal{M}(x) ), \View ( B\leftrightarrow \mathcal{M}(x') ))\preceq d.$$ We have 
	\begin{align}
		&\sup_B D(\View( B\leftrightarrow \mathcal{M}(x) ), \View( B\leftrightarrow \mathcal{M}(x') )) \notag  \\
		=&\sup_B(\mathrm{ChainRule}_{A_1,A'_1}((D(\View(B_{a_1} \leftrightarrow \mathcal{M}_{a_1}(x)), \View(B_{a_1} \leftrightarrow \mathcal{M}_{a_1}(x')) ))_{a_1 \in \supp(A_1) \cap \supp(A'_1)}))  \label{eq.chainrule}  \\
		=&\sup_{ (B_{a_1})_{a_1 \in \supp(A_1) \cap \supp(A'_1)} } (\mathrm{ChainRule}_{A_1,A'_1}((D(\View(B_{a_1} \leftrightarrow \mathcal{M}_{a_1}(x)), \View(B_{a_1} \leftrightarrow \mathcal{M}_{a_1}(x')) ))_{a_1 \in \supp(A_1) \cap \supp(A'_1)})) \label{eq.productadv}\\
		=&\mathrm{ChainRule}_{A_1,A'_1}(\sup_B (D(\View(B_{a_1} \leftrightarrow \mathcal{M}_{a_1}(x)), \View(B_{a_1} \leftrightarrow \mathcal{M}_{a_1}(x'))) )_{a_1 \in \supp(A_1) \cap \supp(A'_1)} ), \label{eq.cont}
	\end{align}
	where \eqref{eq.chainrule} follows from the chain rule, and \eqref{eq.productadv} is because that in the case that the first message $a_1$ comes from the mechanism, specifying an adversary $B$ for the entire mechanism is equivalent to specifying $B_{a_1}$ for ever $a_1$, then the set of deterministic adversary strategies $B$ we need to sup over is a product set over adversary strategies $B_{a_1}$. Equation \eqref{eq.cont} follows from that the $\mathrm{ChainRule}_{A_1,A'_1}$ function is continuous with respect to each variable. 
	For every $a_1\in \supp(A_1) \cap \supp(A'_1)$, define $d_{a_1}=\sup_{B_{a_1}}D(\View(B_{a_1} \leftrightarrow \mathcal{M}_{a_1}(x)), \View(B_{a_1} \leftrightarrow \mathcal{M}_{a_1}(x')) )$.
	Then $\mathcal{M}_{a_1}$ is $d_{a_1}$-$\mathcal{D}$ DP, by induction, there exists a pair of random variables $Y_{a_1}, Y'_{a_1}$ and a post-processing $\mathcal{P}_{a_1}$ such that 
	$$D(Y_{a_1}, Y'_{a_1}) \preceq d_{a_1},$$
	and 
	$$\mathcal{P}_{a_1}(Y_{a_1})\equiv \mathcal{M}_{a_1}(x) \text{ and } \mathcal{P}_{a_1}(Y'_{a_1})\equiv \mathcal{M}_{a_1}(x').$$
	Let $Y_{A_1}$ be the random variable that defined as $Y_{A_1}|_{A_1=a_1}\sim Y_{a_1}$. $Y'_{A_1}$ is defined similarly.
    By the chain rule, we have 
	\begin{align*}
		\MoveEqLeft D((A_1,Y_{A_1}), (A'_1,Y'_{A_1})) \\
		&= \mathrm{ChainRule}_{A_1,A'_1}( (D(Y_{a_1}, Y'_{a_1}))_{a_1 \in \supp(A_1) \cap \supp(A'_1)}) \\
		&\preceq \mathrm{ChainRule}_{A_1,A'_1}( (d_{a_1})_{a_1 \in \supp(A_1) \cap \supp(A'_1)}) \tag{by monotonicity of $\mathrm{ChainRule}$} \\
		&=\mathrm{ChainRule}_{A_1,A'_1}( (\sup_{B_{a_1}}D(\View(B_{a_1} \leftrightarrow \mathcal{M}_{a_1}(x)), \View(B_{a_1} \leftrightarrow \mathcal{M}_{a_1}(x')) ))_{a_1 \in \supp(A_1) \cap \supp(A'_1)} )\\
		&=\mathrm{ChainRule}_{A_1,A'_1}(\sup_B (D(\View(B_{a_1} \leftrightarrow \mathcal{M}_{a_1}(x)), \View(B_{a_1} \leftrightarrow \mathcal{M}_{a_1}(x'))) )_{a_1 \in \supp(A_1) \cap \supp(A'_1)} )\\
		&=\sup_{ B} (\mathrm{ChainRule}_{A_1,A'_1}((D(\View(B_{a_1} \leftrightarrow \mathcal{M}_{a_1}(x)), \View(B_{a_1} \leftrightarrow \mathcal{M}_{a_1}(x')) ))_{a_1 \in \supp(A_1) \cap \supp(A'_1)}))\\
		&\preceq d
	\end{align*}
	Let $Y=(A_1,Y_{A_1})$ and $Y'=(A'_1,Y'_{A_1})$, we define the post-processing $\mathcal{P}$ as 
	$$\mathcal{P}((a_1,y), q_1, \ldots, q_{m-1})=(a_1, \mathcal{P}_{a_1}(y,  q_1, \ldots, q_{m-1})).$$

\end{proof}

We now use Theorem \ref{thm.post_gen} to prove that
the concurrent composition of interactive mechanisms can be reduced to the composition of the non-interactive mechanisms. 

\begin{theorem}[Theorem \ref{thm.fdp} generalized]\label{thm.concomp_d}
	Suppose that the generalized probability distance $(\mathcal{D}, \preceq, D)$ satisfies the chain rule, every $d\in \mathcal{D}$ satisfies the coupling property, and $(\mathcal{D}, \preceq)$ is complete. Suppose for all non-interactive mechanism $\mathcal{M}_1, \ldots, \mathcal{M}_k$ such that $\mathcal{M}_i$ is $d_i$-$\mathcal{D}$ DP for $i=1,2 \ldots, k$, their composition $\comp(\mathcal{M}_1, \ldots, \mathcal{M}_k)$ is $d$-$\mathcal{D}$ DP, then the concurrent composition $\concomp(\mathcal{M}_1, \ldots, \mathcal{M}_k)$  of interactive mechanisms $\mathcal{M}_1, \ldots, \mathcal{M}_k$ such that $\mathcal{M}_i$ is $d_i$-$\mathcal{D}$ DP is also $d$-$\mathcal{D}$ DP.
\end{theorem}

\begin{proof}[Proof of Theorem \ref{thm.concomp_d}]

	Following Theorem \ref{thm.post_gen}, for every interactive $d_j$-$\mathcal{D}$ DP mechanism $\mathcal{M}_j$, $j=1,\ldots, k$, and every pair of neighboring datasets $x,x'$, there exists a pair of of random variables $Y_j,Y'_j$ and an interactive post-processing $\mathcal{P}_j$ such that $D(Y_j, Y'_j)\preceq d_j$, and for every adversary $B\in \mathcal{B}$, $\View(B\leftrightarrow \mathcal{M}_j(x))$ (resp.,$\View(B\leftrightarrow \mathcal{M}_j(x'))$ ) is identically distributed as $\View(B\leftrightarrow \mathcal{P}_j(Y_j))$ (resp., $\View(B\leftrightarrow \mathcal{P}_j(Y'_j))$). Since $Y_j,Y'_j$, $j=1,\ldots, k$, are noninteractive random variables, which can be viewed as the output distributions of a noninteractive mechanism $\mathcal{N}_j$ on $x,x'$. Suppose $\comp(\mathcal{N}_1, \ldots, \mathcal{N}_k)$ is $d$-$\mathcal{D}$ DP. By the post-processing property, we know that $\comp((\mathcal{P}_1(\mathcal{N}_1), \ldots,\mathcal{P}_k(\mathcal{N}_k))$ is also $d$-$\mathcal{D}$ DP. Therefore, we have that $\concomp(\mathcal{M}_1, \ldots, \mathcal{M}_k)$ is also $d$-$\mathcal{D}$ DP.
	
\end{proof}

\section{Concurrent Composition of R\'enyi DP}\label{sec.RDP}

In this section, we give a different and simpler proof of the optimal concurrent composition of R\'enyi DP given in \cite{lyu2022composition}:

\begin{theorem}[\cite{lyu2022composition}]\label{thm.rdp2}
	For all $\alpha>1$, $k\in \mathbb{N}$, $\eps_1,\ldots, \eps_k>0$, and all interactive mechanisms $\mathcal{M}_1, \ldots, \mathcal{M}_k$ such that $\mathcal{M}_i$ is $(\alpha,\eps_i)$-RDP for $i=1,2 \ldots, k$, 
	the concurrent composition $\concomp(\mathcal{M}_1, \ldots, \mathcal{M}_k)$ of interactive mechanisms $\mathcal{M}_1, \ldots, \mathcal{M}_k$ is $(\alpha,\sum_{i=1}^k\eps_i)$-RDP.
\end{theorem}

We prove this theorem by characterizing optimal $\alpha$-RDP adversary strategy in Lemma \ref{lemma.optadv}. 

\begin{definition}[Optimal $\alpha$-RDP adversary]
	For an interactive mechanism $\mathcal{M}$, and neighboring datasets $x$ and $x'$, an optimal $\alpha$-RDP adversary with respect to $x$ and $x'$ is a strategy $B^{\mathrm{OPT}}$ such that for all adversary strategies $B$, 
	\begin{equation*}
		D_\alpha( \View(B^{\mathrm{OPT}}\leftrightarrow \mathcal{M}(x)) || \View(B^{\mathrm{OPT}}\leftrightarrow \mathcal{M}(x'))   )\ge D_\alpha( \View(B\leftrightarrow \mathcal{M}(x)) || \View(B\leftrightarrow \mathcal{M}(x'))   ).
	\end{equation*}
\end{definition}

We show that the optimal adversary strategy against the concurrent composition of $k$ mechanisms can be decomposed as a product of optimal adversaries against each mechanism independently: 

\begin{lemma}\label{lemma.optconcomp}
	Let $B^{\mathrm{OPT} (1)}, B^{\mathrm{OPT} (2)}, \ldots, B^{\mathrm{OPT} (k)}$ be optimal $\alpha$-RDP adversaries against $\mathcal{M}_1, \mathcal{M}_2, \ldots, \mathcal{M}_k$. Then 
	\begin{equation*}
		B^{\mathrm{OPT}} =B^{\mathrm{OPT} (1)} \times B^{\mathrm{OPT} (2)} \times \ldots \times B^{\mathrm{OPT} (k)}
	\end{equation*}
is an optimal $\alpha$-RDP adversary against $\concomp(\mathcal{M}_1, \mathcal{M}_2, \ldots, \mathcal{M}_k)$, where $B^{\mathrm{OPT} (1)} \times B^{\mathrm{OPT} (2)}$ denotes the adversary's strategy where it takes $B^{\mathrm{OPT} (1)}$ to interact with $\mathcal{M}_1$ and takes $B^{\mathrm{OPT} (2)}$ to interact with $\mathcal{M}_2$.
\end{lemma}

Although this property of the optimal adversary strategy can be derived as a consequence of the optimal concurrent composition of R\'enyi DP in \cite{lyu2022composition}, we take a different approach to first prove this property and then use it to prove the optimal concurrent composition theorem for R\'enyi DP. 

Our proof relies on the following two properties of R\'enyi divergence: the monotonicity property in Lemma \ref{lemma.monotone} and the independence property in Lemma \ref{lemma.indep}. The R\'enyi divergence is defined as follows.
\begin{definition}[R\'enyi divergence \cite{renyi1961measures}]
	For two probability distributions $P$ and $Q$, the R\'enyi divergence of order $\alpha>1$ is 
	\begin{equation*}
		D_\alpha(P||Q)=\frac{1}{\alpha-1}\log \mathrm{E}_{x\sim Q} \left[\frac{P(x)}{Q(x)}\right]^\alpha.
	\end{equation*}
\end{definition}

\begin{lemma}\label{lemma.monotone}
	For any two tuples of jointly distributed random variables $(U,V,W)$ and $(U',V',W')$ over the same measureable space,  if for every $u\in \supp(U)$, we have 
	\begin{equation*}
		D_\alpha(V|_{U=u}||V'|_{U'=u}) \le D_\alpha(W|_{U=u}||W'|_{U'=u}),
	\end{equation*}
then 
	\begin{equation*}
	D_\alpha((U, V)||(U',V')) \le D_\alpha((U,W)||(U',W')).
\end{equation*}
\end{lemma}

\begin{proof}
	\begin{align*}
	\MoveEqLeft D_\alpha((U, V)||(U',V')) \\
	&=\frac{1}{\alpha-1}\log \sum_{u,v}\frac{ \left( \Pr(U=u)\Pr(V=v|U=u) \right)^\alpha }{ \left( Pr(U'=u)\Pr(V'=v|U'=u) \right)^{\alpha -1} } \\
	&=\frac{1}{\alpha-1}\log \sum_{u}\frac{ \left( \Pr(U=u) \right)^\alpha} { \left( Pr(U'=u)\right)^{\alpha -1} }\left( \sum_{v}\frac{ \left( \Pr(V=v|U=u) \right)^\alpha} { \left( Pr(V'=v|U'=u)\right)^{\alpha -1} }\right) \\
	&=\frac{1}{\alpha-1}\log \sum_{u}\frac{ \left( \Pr(U=u) \right)^\alpha} { \left( Pr(U'=u)\right)^{\alpha -1} }\left( \exp(\alpha-1) D_\alpha(V|_{U=u}||V'|_{U'=u}) \right).
	\end{align*}
Hence, $D_\alpha((U, V)||(U',V')) $ is monotonically increasing if $D_\alpha(V|_{U=u}||V'|_{U'=u})$ increases. So if $D_\alpha(V|_{U=u}||V'|_{U'=u}) \le D_\alpha(W|_{U=u}||W'|_{U'=u})$, then we have $D_\alpha((U, V)||(U',V')) \le D_\alpha((U,W)||(U',W')).$

\end{proof}

\begin{lemma}\label{lemma.indep}
	For any two pairs of random variables $U, U'$ and $V, V'$, if $U$ and $V$ ($U'$ and $V'$ resp.) are independent, then 
	\begin{equation*}
		D_\alpha((U, V)||(U',V')) =D_\alpha(U||U') + D_\alpha(V||V').
	\end{equation*}
\end{lemma}

\begin{proof}
	\begin{align*}
	\MoveEqLeft	D_\alpha((U, V)||(U',V')) \\
	&=\frac{1}{\alpha-1}\log \sum_{u,v}\frac{ \left( \Pr(U=u)\Pr(V=v) \right)^\alpha }{ \left( Pr(U'=u)\Pr(V'=v) \right)^{\alpha -1} } \\
	&=\frac{1}{\alpha-1}\log \sum_{u}\frac{ \left( \Pr(U=u) \right)^\alpha} { \left( Pr(U'=u)\right)^{\alpha -1} }\left( \exp(\alpha-1) D_\alpha(V||V') \right) \\
	&=D_\alpha(U||U') + D_\alpha(V||V').
	\end{align*}
\end{proof}

The following lemma describes the optimal adversary's strategy against an interactive mechanism. 
 The proof of Lemma \ref{lemma.optadv} uses the monotonicity property of R\'enyi divergence.

\begin{lemma}\label{lemma.optadv}
	The optimal adversary $B^{\mathrm{OPT}}$ with respect to $x$ and $x'$ , which is the adversary strategy that maximizes the R\'enyi divergence of the views for all fixed $x,x'$,
	chooses the first query $q_1$ to maximize 
\begin{equation*}
D_\alpha\left(  \left(A_1, \View\left(   B^{\mathrm{OPT}}_{q_1, A_1} \leftrightarrow M_{q_1, A_1}(x)  \right)  \right)  ||  \left(A'_1, \View \left(  B^{\mathrm{OPT}}_{q_1, A'_1} \leftrightarrow M_{q_1, A'_1}(x')  \right)  \right)    \right) ,
\end{equation*}
where $A_1= \mathcal{M}(x,q_1)$ and $A'_1= \mathcal{M}(x',q_1)$, and $B^{\mathrm{OPT}}_{q_1, a_1}$ is any optimal adversary against $\mathcal{M}_{q_1, a_1}$, which denotes the subsequent mechanism when fixing $q_1, a_1$.
\end{lemma}

\begin{proof}
We decompose the view of the adversary into two parts: the first answer $A_1$ to the query $q_1$, and the view of the subsequent interaction. Fixing $q_1$, for every adversary $B$, we have
\begin{align}
	\MoveEqLeft D_\alpha( \View(B\leftrightarrow \mathcal{M}(x)) || \View(B\leftrightarrow \mathcal{M}(x'))   ) \label{eq.divergence} \\
&=D_\alpha\left(  \left(A_1(q_1), \View\left(   B_{q_1, A_1} \leftrightarrow \mathcal{M}_{q_1, A_1}(x)  \right)  \right)  ||  \left(A'_1(q_1), \View \left(  B_{q_1, A'_1} \leftrightarrow \mathcal{M}_{q_1, A'_1}(x')  \right)  \right)    \right) , \notag
\end{align}
where $B_{q_1, A_1} \leftrightarrow \mathcal{M}_{q_1, A_1}$ denotes the subsequent interaction. For every $a_1\in \supp(A_1)\cap\supp(A'_1)$ and for every $B_{q_1, A_1}$, by the definition of $B^{\mathrm{OPT}}_{q_1, A_1}$, we have 
\begin{align*}
	\MoveEqLeft   D_\alpha\left(  \View(B_{q_1, A_1}\leftrightarrow \mathcal{M}_{q_1,A_1}(x))|_{A_1=a_1} || \View(B_{q_1,A'_1}\leftrightarrow \mathcal{M}_{q_1,A'_1}(x'))|_{A'_1=a_1}   \right)  \\
	&\le D_\alpha \left( \View(B^{\mathrm{OPT}}_{q_1, A_1}\leftrightarrow \mathcal{M}_{q_1,A_1}(x))|_{A_1=a_1} || \View(B^{\mathrm{OPT}}_{q_1,A'_1}\leftrightarrow \mathcal{M}_{q_1,A'_1}(x'))|_{A'_1=a_1}   \right).
\end{align*}
Thus, we have 
\begin{align}
	\MoveEqLeft  D_\alpha\left(  \left(A_1(q_1), \View\left(   B_{q_1, A_1} \leftrightarrow \mathcal{M}_{q_1, A_1}(x)  \right)  \right)  ||  \left(A'_1(q_1), \View \left(  B_{q_1, A'_1} \leftrightarrow \mathcal{M}_{q_1, A'_1}(x')  \right)  \right)    \right) \notag \\
&\le D_\alpha\left(  \left(A_1(q_1), \View\left(   B^{\mathrm{OPT}}_{q_1, A_1} \leftrightarrow \mathcal{M}_{q_1, A_1}(x)  \right)  \right)  ||  \left(A'_1(q_1), \View \left(  B^{\mathrm{OPT}}_{q_1, A'_1} \leftrightarrow \mathcal{M}_{q_1, A'_1}(x')  \right)  \right)    \right), \label{eq.optB}
\end{align}
where \eqref{eq.optB} follows from Lemma \ref{lemma.monotone}. It implies that in order to maxmize \eqref{eq.divergence}, it suffices to choose $q_1$ to maximize the quantity in \eqref{eq.optB}.
\end{proof}

We then use Lemma \ref{lemma.optadv} to prove Lemma \ref{lemma.optconcomp}.


\begin{proof}[proof of Lemma \ref{lemma.optconcomp}.]
	We will use induction on the rounds of messages to prove this lemma. We can use induction argument because of the assumption of finite communication. Without loss of generality, suppose the first query from the adversary is sent to $\mathcal{M}_1$, and we use $\concomp = \concomp(\mathcal{M}_1, \mathcal{M}_2, \ldots, \mathcal{M}_k)$ to simplify the notation.
	Following Lemma \ref{lemma.optadv}, the optimal adversary $B^{\mathrm{OPT}}$ chooses $q_1$ as follows.
	\begin{equation*}
	B^{\mathrm{OPT}}=\argmax_{q_1}  D_\alpha\left(  \left(A_1, \View\left(   B^{\mathrm{OPT}}_{q_1, A_1} \leftrightarrow \concomp_{q_1, A_1}(x)  \right)  \right)  ||  \left(A'_1, \View \left(  B^{\mathrm{OPT}}_{q_1, A'_1} \leftrightarrow \concomp_{q_1, A'_1}(x')  \right)  \right)    \right),
	\end{equation*}
where $A_1=\mathcal{M}_1(x,q_1)$ and $A'_1=\mathcal{M}_1(x',q_1)$.
By induction, we assume that $B^{\mathrm{OPT}}_{q_1, A_1}= B^{\mathrm{OPT} (1)}_{q_1, A_1} \times  B^{\mathrm{OPT} (2)} \times \ldots \times B^{\mathrm{OPT} (k)} $. Let $V_{q_1, A_1}=\View\left( B^{\mathrm{OPT} (1)}_{q_1, A_1} \leftrightarrow \mathcal{M}_{q_1, A_1}(x)  \right) $, and similarly, $V'_{q_1, A_1}=\View\left( B^{\mathrm{OPT} (1)}_{q_1, A'_1} \leftrightarrow \mathcal{M}_{q_1, A'_1}(x')  \right) $. Let $\tilde{V}=\View \left(  B^{\mathrm{OPT} (2)} \times \ldots \times B^{\mathrm{OPT} (k)} \leftrightarrow \concomp(\mathcal{M}_2, \ldots, \mathcal{M}_k) (x) \right) $ and \\ $\tilde{V}'=\View \left(  B^{\mathrm{OPT} (2)} \times \ldots \times B^{\mathrm{OPT} (k)} \leftrightarrow \concomp(\mathcal{M}_2, \ldots, \mathcal{M}_k)(x') \right) $. With these notations, we have
\begin{align}
	\MoveEqLeft \argmax_{q_1}  D_\alpha\left(  \left(A_1, \View\left(   B^{\mathrm{OPT}}_{q_1, A_1} \leftrightarrow \concomp_{q_1, A_1}(x)  \right)  \right)  ||  \left(A'_1, \View \left(  B^{\mathrm{OPT}}_{q_1, A'_1} \leftrightarrow \concomp_{q_1, A'_1}(x')  \right)  \right)    \right) \notag \\
	=& \argmax_{q_1} D_\alpha\left(  \left(  A_1, V_{q_1, A_1}, \tilde{V}  \right) || \left(  A'_1, V'_{q_1, A_1}, \tilde{V}'  \right)    \right) \notag  \\
	=& \argmax_{q_1} \left\lbrace  D_\alpha\left(  \left(  A_1, V_{q_1, A_1} \right) || \left(  A'_1, V'_{q_1, A_1} \right)    \right) + D_\alpha(\tilde{V}  ||  \tilde{V}' )  \right\rbrace , \tag{by Lemma \ref{lemma.indep}} \\
	=& \argmax_{q_1}  D_\alpha\left(  \left(  A_1, V_{q_1, A_1} \right) || \left(  A'_1, V'_{q_1, A_1} \right)    \right) \notag \\
	=&B^{\mathrm{OPT} (1)}.
\end{align}
Therefore, the optimal adversary $B^{\mathrm{OPT}}$ chooses $q_1$  just as the optimal adversary $B^{\mathrm{OPT}(1)}$ against only $M_1$. Since $B^{\mathrm{OPT}}_{q_1, A_1}= B^{\mathrm{OPT} (1)}_{q_1, A_1} \times  B^{\mathrm{OPT} (2)} \times \ldots \times B^{\mathrm{OPT} (k)} $, we have 
$B^{\mathrm{OPT}}=B^{\mathrm{OPT} (1)} \times  B^{\mathrm{OPT} (2)} \times \ldots \times B^{\mathrm{OPT} (k)} $, completing the proof.
\end{proof}

We now prove Theorem \ref{thm.rdp2} using Lemma \ref{lemma.optconcomp}.

\begin{proof}[Proof of Theorem \ref{thm.rdp2}]
Following Lemma \ref{lemma.optconcomp}, we have 
\begin{align*}
	\MoveEqLeft D_\alpha \left( \View \left( B^{\mathrm{OPT}} \leftrightarrow \concomp (x) \right)  ||  \View \left( B^{\mathrm{OPT}} \leftrightarrow \concomp (x') \right)   \right) \\
	=& D_\alpha \left( \View \left( B^{\mathrm{OPT}(1)} \leftrightarrow \mathcal{M}_1 (x) \right)  ||  \View \left( B^{\mathrm{OPT}(1)} \leftrightarrow \mathcal{M}_1(x') \right)   \right)+ \ldots \\
	&+ D_\alpha \left( \View \left( B^{\mathrm{OPT}(k)} \leftrightarrow \mathcal{M}_k (x) \right)  ||  \View \left( B^{\mathrm{OPT}(k)} \leftrightarrow \mathcal{M}_k(x') \right)   \right) \\
	=&\sum_i^k\eps_i,
\end{align*}
completing the proof.
\end{proof}

\section*{Acknowledgments}
We are grateful to Xin Lyu for pointing out the error in the proof of our previously claimed concurrent composition theorem for R\'enyi DP~\cite{lyu2022composition}. We also thank an anonymous reviewer for pointing out the error in our previous proof of continuity.

	\bibliography{ref}
	\bibliographystyle{alpha}

\end{document}